\pdfoutput=1

\documentclass[conference]{IEEEtran}

\ifCLASSINFOpdf
\usepackage[pdftex]{graphicx}
\else
\fi

\usepackage{pbox}
\usepackage{cite}
\usepackage{graphicx}
\usepackage{float}
\usepackage{amsmath}
\usepackage[utf8]{inputenc}
\usepackage[final]{pdfpages}
\usepackage{pdfpages}
\usepackage{lipsum}
\usepackage{textcase}
\usepackage{url}
\usepackage{amsmath,esint}
\usepackage{longtable}
\usepackage{array}
\usepackage{float}
\usepackage{caption}
\usepackage{tabulary}
\usepackage[section]{placeins}
\usepackage{hyperref}
\usepackage{subfigure}

\restylefloat{table}
\usepackage[normalem]{ulem} 

\newcolumntype{P}[1]{>{\centering\arraybackslash}p{#1}}

	\newcommand\blfootnote[1]{%
		\begingroup
		\renewcommand\thefootnote{}\footnote{#1}%
		\addtocounter{footnote}{-1}%
		\endgroup
	}

\begin{document}
\title{A Survey of Air-to-Ground Propagation Channel Modeling for Unmanned Aerial Vehicles}

\author{\IEEEauthorblockN{Wahab Khawaja\IEEEauthorrefmark{1},~Ismail Guvenc\IEEEauthorrefmark{1},~David W. Matolak\IEEEauthorrefmark{2}, Uwe-Carsten Fiebig\IEEEauthorrefmark{3}, Nicolas Schneckenberger\IEEEauthorrefmark{3}}

\IEEEauthorblockA{\IEEEauthorrefmark{1}Department of Electrical and Computer Engineering, North Carolina State University, Raleigh, NC}

\IEEEauthorblockA{\IEEEauthorrefmark{2}Department of Electrical Engineering, University of South Carolina, Columbia, SC}

\IEEEauthorblockA{\IEEEauthorrefmark{3}German Aerospace Center (DLR) Institute of Communications and Navigation, Wessling, Germany}


Email: \{wkhawaj, iguvenc\}@ncsu.edu, \{matolak\}@cec.sc.edu, \{uwe.fiebig, nicolas.schneckenburger\}@dlr.de
	
}

\maketitle

\blfootnote{This work has been supported in part by NSF under the grant CNS-$1453678$ and by NASA under the Federal Award ID number
NNX17AJ94A.}

\begin{abstract}
In recent years, there has been a dramatic increase in the use of unmanned aerial vehicles (UAVs), particularly for small UAVs, due to their affordable prices, ease of availability, and ease of operability. Existing 
and future applications 
of UAVs include 
remote surveillance and monitoring, relief operations, package delivery, and communication backhaul infrastructure. 
Additionally, UAVs are envisioned as an important component of 5G wireless  technology and beyond. The unique application scenarios for UAVs necessitate accurate air-to-ground (AG) propagation channel models for designing and evaluating UAV communication links for control/non-payload as well as payload data transmissions. These AG propagation models have not been investigated in detail when compared to terrestrial propagation models. In this paper, a comprehensive survey is provided on available AG channel measurement campaigns, large and small scale fading channel models, their limitations, and future research directions for UAV communication scenarios.       


\end{abstract}

\IEEEpeerreviewmaketitle

\section{Introduction}\label{Section:Introduction}
Use of commercial unmanned aerial vehicles (UAVs) has recently seen exceptional growth that is forecast to continue in the near future. The benefits of easy operability, multiple flight controls, high maneuverability, and increasing payload weight of currently available UAVs have led to their introduction into many real time civilian applications including remote surveillance, filming, disaster relief, goods transport, and communication relaying, not to mention recreation. According to statistics provided by the market research company Tractica, the shipment of commercial UAVs units is expected to reach $2.7$~million in $2025$ with the services offered rising to \$$8.7$~billion in the next decade\cite{Tractica}.

UAVs are also termed unmanned aerial systems (UAS), and commonly known by the term “drones.” These aircraft can vary in size from small toys that fit in the palm of a human hand (where the “unmanned” designation is unnecessary) to large military aircraft such as the General Atomics MQ-9 Reaper (commonly termed Predator) \cite{Drones}, with a wingspan over 15 meters. The small, battery powered toys generally can fly for up to $15$ minutes, whereas the larger UAVs are designed for long-endurance ($30$ hours), high-altitude operations (higher than $15$ km). 

In this paper our focus is on the smaller UAVs. Various organizations have developed classifications for UAVs according to size, with designations large, medium, and small being typical. In the US, the Federal Aviation Administration (FAA) has issued rules for small UAVs weighing less than 55 pounds ($25$ kg) \cite{FAA_new}. Highlights of these rules include the requirement for a visual line-of-sight (LOS) from pilot to aircraft, flight under daylight or during twilight (within $30$ minutes of official sunrise/sunset) with appropriate lighting for collision avoidance, a maximum flight ceiling of $400$ feet ($122$~m) above the ground (higher if the UAV is within $122$~m of a construction site), and a maximum speed of $100$ mph ($87$ knots, or $161$ km/h). Restrictions also apply regarding proximity to airports, and generally, a licensed pilot must operate or supervise UAV operation.

One of the promising applications of UAVs is in supporting broadband wireless cellular communications in hot spot areas during peak demand events and in cases of a natural calamity where the existing communication infrastructure is damaged. It is expected that future 5G implementation will include UAVs as autonomous communicating nodes for providing low latency and highly reliable communications, at least in some situations. Qualcomm is testing the operability of UAVs for current LTE and future 5G cellular applications \cite{Qual}. In addition, UAVs can act as mobile wireless access points in different network topologies supporting different protocols of IEEE $802.11$. Facebook and Google are also exploring the possibility of using UAVs for Internet connectivity to remote areas using UAVs \cite{CNN}. 

Air-to-ground~(AG) communications can be traced back to $1920$ \cite{history1}, with manually operated radio telegraphs. Lower and medium frequency bands were used in the early $1930s$ but did not support simultaneous voice communications in both directions (AG and ground-to-air (GA)). From the early $1940s$, double sideband amplitude modulation~(DSB-AM) in the frequency band ($118$~MHz - $137$~MHz) lying in the very high frequency~(VHF) band was adopted for voice communications between pilots and ground controllers. This system supported a maximum of 140 channels until 1979. Multiplexing and multiple access were frequency division with \textit{manual} channel assignment by air traffic control. In more dense air traffic spaces, to enable larger numbers of simultaneous transmissions, 25~kHz DSB-AM channels were subdivided into three channels of width 8.33~kHz. The civilian aeronautical AG communications continues to use the reliable analog DSB-AM system today, although since $1990$ some small segments of the VHF band in some geographic locations are being upgraded to a digital VHF data link that can in principle support  2280 channels \cite{history3,manned_2}. This system employs time-division as well as frequency-division, with single-carrier phase-shift keying modulation. Military AG communications uses different frequency bands (ultra-high frequency) and modulation schemes for short and long ranges\cite{militaryAG}. Due to very low data rates, the civil aviation systems cannot support modern AG communication requirements. In $2007$, use of portion of the L-band was suggested for new civil aviation systems, and two such systems known as L-band Digital Aeronautical Communications Systems, or LDACS, were developed\cite{manned_2}. Due to compatibility with numerous existing systems that operate in the L-band, the LDACS system is still being refined. LDACS is currently being standardized by the International Civil Aviation Organization (ICAO).

There are numerous studies available in the literature on the characteristics of aeronautical channels \cite{manned_1,history3,aeronautical1,aeronautical2,Survey_Matolak}. Aeronautical communications can be broadly classified into communications between the pilot or crew with the ground controller and wireless data communication for passengers. Both of these types of communication are dependent on the flight route characteristics. In \cite{manned_1} the propagation channel is divided into three main phases of flight, termed as parking and taxiing, en-route, and take off and landing. Each phase of flight was described by different channel characteristics~(type of fading, Doppler spread, and delay), but this relatively early paper was not comprehensive nor fully supported by measurements.

There are also long distance AG propagation channel studies available for satellites and high altitude platforms (HAPs). The AG propagation channel in these studies can be considered as a UAV communication channel, but due to long distances from the earth surface, normally greater than $17$~km, modeling of these links may also need to take into account upper atmospheric effects. Depending on frequency and UAV altitude, they may also be much more susceptible to lower tropospheric effects such as fading from hydrometeors~\cite{HAPs_satellites}. For most of these longer distance platforms, a LOS component is required because of power limitations, hence the AG channel amplitude fading is typically modeled as Ricean~\cite{HAP1}. As the deployment of UAVs as communication nodes in the near future is expected to be at much lower altitudes compared to that of HAPs and satellites, in this survey we focus only on lower altitude UAV AG propagation channels.

In order to fulfill the ever increasing demands of high rate data transfer in the future using UAVs in different environments, robust and accurate AG propagation channel models are required. The available AG propagation channel models used for higher altitude aeronautical communications generally cannot be employed directly for low-altitude UAV communications. Small UAVs may also possess distinct structural and flight characteristics such as different airframe shadowing features due to unique body shapes and materials, and potentially sharper pitch, roll, and yaw rates of change during flight. The AG channel for UAVs has not been studied as extensively as the terrestrial channel.

The available UAV based AG wireless propagation channel research can be largely categorized into two major portions. The first one is payload communications, where the payload can be narrow-band or wide-band and is mostly application dependent. The second one is control and non-payload communications (CNPC) for telemetric control of UAVs. Most payload UAV communications employs the unlicensed bands e.g., $900$~MHz, $2.4$~GHz, and $5.8$~GHz; this is not preferred by the aviation community as these bands can be congested and may be easily jammed. In the USA, CNPC is potentially planned for a portion of L-band ($0.9$~GHz - $1.2$~GHz) and C-band ($5.03$~GHz - $5.091$~GHz), although as is common in spectrum allocation, use of these bands is still being negotiated~\cite{CNPC1,CNPC2}. Channel measurements and modeling for UAVs are~(other than bandwidth and carrier frequency) largely independent of whether signaling is for payload or CNPC.

In this study, we will discuss recent channel measurement campaigns and modeling efforts to characterize the AG channel for UAVs. We also describe future research challenges and possible enhancements. To the best of our knowledge, there is to date no comprehensive survey on AG propagation channel models for UAV  wireless communications.

The rest of the paper is organized as follows. Section~II explains the UAV AG propagation channel characteristics. The AG channel measurements and associated features are described in Section~III, and Section~IV discusses AG propagation channel models, including models based on ray tracing simulations. Future challenges and research directions are provided in Section~V, and concluding remarks follow in Section~VI. All acronyms and variables used throughout the paper are given in Table~\ref{Table:Table_acronym} and Table~\ref{Table:Table_variables}, respectively.

\section{UAV AG Propagation Channel Characteristics}
\begin{figure*}[!t]
	\centering
	\includegraphics[width=1.6\columnwidth]{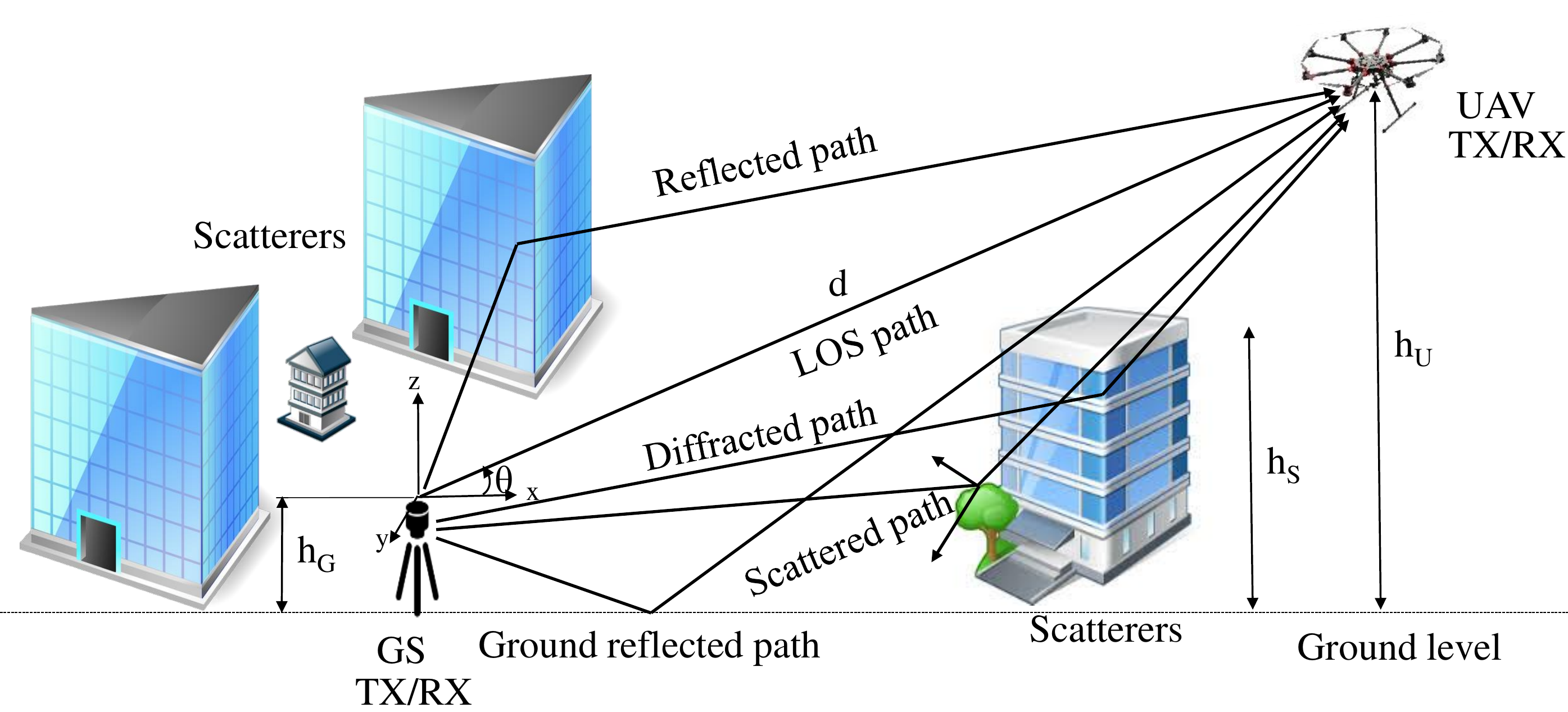}
	\caption{A typical air-to-ground propagation scenario with a UAV.}\label{Fig:General_fig}
\end{figure*}

In this section, salient characteristics of UAV AG propagation channels are described. A common AG propagation scenario is shown in Fig.~\ref{Fig:General_fig} in the presence of terrestrial obstacles which are also commonly referred as \textit{scatterers}. In the figure, $h_{\rm G}$, $h_{\rm S}$, $h_{\rm U}$ represents the height of the ground station~(GS), scatterers, and UAV above the ground, respectively, $d$ is the slant range between the UAV antennas and the GS, and $\theta$ is the elevation angle between GS and UAV antennas. (We note that airborne scatterers may be present as well, but for this paper, for the AG link, we neglect this secondary condition.)

\subsection{Comparison of UAV AG Propagation with Terrestrial}
The AG channel exhibits distinctly different characteristics from those of other well studied terrestrial communication channels, e.g., the urban channel. On the one hand, there is the inherent advantage over terrestrial communications in terms of a higher likelihood of LOS propagation.This reduces transmit power requirements and can translate to higher link reliability as well. In cases where only non-LOS~(NLOS) paths exist, when the elevation angle to the UAV is large enough, the AG channel may incur smaller diffraction and shadowing losses than near ground terrestrial links.

On the other hand, the AG channel can exhibit significantly higher rates of change than typical terrestrial communication channels because of UAV velocities. When the channel is modeled statistically, this can mean that the channel's statistics are approximately constant (the channel is wide-sense stationary) for only a small spatial extent. This is often loosely termed "non-stationarity." If the UAV is not in the direct vicinity of scattering objects or the GS, the characteristics of the channel could instead actually change very slowly, especially for hovering UAVs. In such a case, adverse propagation conditions, e.g., deep fades of the received signal, may last several seconds or even minutes, hence common communication techniques of interleaving or averaging would not be effective. In many cases, when UAV altitudes are well above scattering objects, the AG channel's "non-stationarity" will be attributable to the direct surroundings of the GS, e.g., the close by buildings or the ground surface composition around the GS.

Additionally, AG communications with UAVs face many other challenges, due to arbitrary mobility patterns and diverse types of communication applications~\cite{UAV_challenges1,UAV_challenges2,UAV_challenges3,UAV_Specs3}. As an aerial node, some of the UAV specifics that need to be taken into account include airframe shadowing, mechanical and electronic noise from UAV electronics and motors, and finally antenna characteristics, including size, orientation, polarization, and array operation (e.g., beam steering) for multiple-input-multiple-output~(MIMO) systems. For UAVs in motion, the effect of Doppler shifts and spread must also be considered for specific communication applications~\cite{UAV_doppler1,UAV_doppler2}. For a given setting, an optimum UAV height may need to be considered, e.g., for maintaining LOS in that environment\cite{UAV_sim3}. 

\begin{table*}[htbp]
	\begin{center}		
    \footnotesize
		\caption{Acronyms used in this paper.}\label{Table:Table_acronym}
        	\begin{tabular}{|P{1.8cm}|P{5.5cm}|P{1.8cm}|P{5.5cm}|}
			\hline
			\textbf{Acronym}&\textbf{Text}&\textbf{Acronym}&\textbf{Text}\\
            \hline
            AA & Air-to-air & AG & Air-to-ground \\
        	\hline 
            AWGN & Additive white Gaussian noise & BER & Bit error rate \\
        	\hline
            BPSK & Binary phase shift keying & BW & Bandwidth \\
            \hline
            CDF & Cumulative distribution function & CFO & Carrier frequency offset  \\
            \hline
            CIR & Channel impulse response & CNPC & Control and non-payload communications \\
        	\hline
            CSI & Channel state information & CTF & Channel transfer function \\
            \hline 
            CW & Continuous wave & DPP & Doppler power profile\\
            \hline
            DS & Doppler spread & DSB-AM & Double sideband amplitude modulation \\
            \hline
            DS-SS& Direct sequence spread spectrum & FAA & Federal aviation administration \\
			\hline
            FMBC & Filter bank multicarrier & FMCW& Frequency modulated continuous wave \\
            \hline
            GA & Ground-to-air & GMSK & Gaussian Minimum Shift Keying \\
           \hline
            GPS & Global positioning system & \\
            \hline 
            GS& Ground station & GSM & Global system for mobile communication \\
			\hline
            HAP & High altitude platform & ICI & Inter-carrier interference \\
        	\hline
            IS-GBSCM & Irregular shaped geometric based stochastic channel model & LAP & Lower altitude platform \\
        	\hline
            LDACS & L-band digital aeronautical communications & LOS & Line-of-sight \\
        	\hline
            LTE & Long term evolution & MIMO & Multiple-input-multiple-output \\
        	\hline
            MISO & Multiple-input-single-output & Mod. Sig. & Modulated signal \\
        	\hline
            MPC & Multipath component & MSK & Minimum shift keying  \\
            \hline
            NGSCM & Non-geometric channel model & NLOS & Non-line-of-sight \\
            \hline
            OFDM & Orthogonal frequency-division multiplexing & OLOS & Obstructed line-of-sight \\
        	\hline
            PAPR & Peak to average power ratio & PDP & Power delay profile \\
            \hline
            PG & Path gain & PL & Path loss \\
            \hline
            PLE & Path loss exponent & PRN & Pseudo-random number \\
        	\hline
            PSD & Power spectral density & RF & Radio frequency \\
        	\hline
            RMS-DS & Root mean square-delay spread & RS-GBSCM & Regular shaped geometric based  stochastic channel model \\
            \hline
            RSS & Received signal strength & RSSI & Received signal strength indicator \\
        	\hline
            RTT & Round trip time & RX & Receiver \\
            \hline
            SDMA & Space-division multiple access & SIMO & Single-input-multiple-output  \\
            \hline
            SISO & Single-input-single-output & SNR & Signal-to-noise-ratio \\
        	\hline
            TDL & Tap-delay-line & TDMA & Time division multiple access \\
            \hline
            TOA & Time-of-arrival & TX & Transmitter \\
        	\hline
            UAS & Unmanned aerial systems & UAV & Unmanned aerial vehicle \\
            \hline
            UMTS & Universal Mobile Telecommunications Service & UWB & Ultra-wideband \\
			\hline
            VHF & Very high frequency  & WSS & Wide sense stationary \\
        	 \hline
            
		  \end{tabular}
		\end{center}
 \end{table*}

\begin{table}[htbp]
	\begin{center}
    \footnotesize
		\caption{Variables used in this paper.}\label{Table:Table_variables}
		\begin{tabular}{|P{1.8cm}|P{4.8cm}|}
			\hline
			\textbf{Acronym}&\textbf{Text}\\
            \hline
            
            $a_{i}$ & Amplitude of $i^{\rm th}$ MPC \\
			\hline
            $c$ & Speed of light \\
            \hline
            $d$&Link distance between TX and RX \\
            \hline
            $d_{0}$&Reference distance between TX and RX \\
            \hline
            $f$ & Frequency instance \\
            \hline
            $f_{\rm c}$ & Carrier frequency \\
            \hline
            $f_{\rm d}^{i}$ & Doppler frequency shift of $i^{\rm th}$ MPC \\
            \hline
            $h_{\rm G}$ & Ground station height\\
            \hline
            $h_{\rm S}$ & Height of scatterer \\
            \hline
            $h_{\rm U}$& UAV altitude above ground \\
            \hline
            $K$-factor & Ricean $K$-factor \\
            \hline
            $M$ & Total number of MPCs \\
			\hline
            $p_{i}(t)$ & MPCs persistence coefficient \\
			\hline
            $P_{\rm R}$& Received power \\
			\hline
            $P_{\rm T}$& Transmit power \\
			\hline	
			$PL_{0}$ & Reference path loss \\
            \hline
            $t$ & Time instance \\
            \hline
            $v$ & Velocity of UAV \\
            \hline
            $v_{\rm max}$ & Maximum speed\\
            \hline
            $\Theta$ & Aggregated phase angles \\
            \hline
            $\gamma$ & Path loss exponent\\
            \hline
            $\tau_{i}$ & Delay of $i^{\rm th}$ MPC \\
			\hline
            $\lambda$ & Wavelength of the radio wave \\
            \hline
			$\phi_{i}$ & Phase of the $i^{\rm th}$ MPC \\
            \hline
			$X$ & Shadowing random variable\\
            \hline
            $\theta$&Elevation angle \\
			\hline	
			$\sigma$ & Standard deviation of shadow fading \\
            \hline
            $\Delta \psi$ & Phase difference between the LOS and ground reflected MPC \\
            \hline
            $\varsigma$ &  Ratio of built up area to total area\\
            \hline
            $\xi$ &  Mean number of buildings per unit area\\
            \hline
            $\Omega$ &  Height distribution of buildings\\
            \hline
            $\alpha$ & Slope of linear least square regression fit\\
            \hline
            $\beta$ & Y-intercept point for the linear least square regression fit \\
            \hline
                        
            \end{tabular}
		      \end{center}
               \end{table}

\subsection{Frequency Bands for UAV AG Propagation}
As with all communication channels, a fundamental consideration is the frequency band, since propagation characteristics can vary significantly with frequency. For the L and C-bands envisioned for CNPC, and for the currently popular unlicensed bands for payload communications, tropospheric attenuations from atmospheric gases and hydrometeors are mostly negligible. This will not be true for operation at higher frequency bands e.g., at Ku, Ka, and other so-called millimeter wave (mmWave) bands, which may be as high as 100 GHz. These higher frequency bands can hence suffer both larger free-space path loss~(PL) as well as tropospheric attenuations. Because of this, these frequency bands will generally be used for short-range AG links.

In contrast to the attenuation characteristics compared with lower frequency bands, mmWave bands offer a large amount of bandwidth, which is their primary appeal for 5G cellular systems. Large bandwidths can be more robust to the larger values of Doppler shift and Doppler spread encountered with UAVs moving at high velocity.

\subsection{Specifics of UAV AG Propagation Channel}\label{Subsection:AG_Prop_channel}
In an AG propagation channel using UAVs, the multipath components (MPCs) appear due to reflections from the earth surface, from terrestrial objects~(ground scatterers), and sometimes from the airframe of the UAV itself. The characteristics of the channel will be dependent on the material, shape, and size of the scattering objects. The strongest MPC apart from the LOS component in an AG propagation scenario is often the single reflection from the earth surface. This gives rise to the well known two ray model.

For high enough frequencies, the scatterers on the ground and around the UAV can be modeled as points scatterers on the surface of two respective cylinders or spheres~\cite{UAV_geometric9,UAV_geometric2} or ellipsoids, and these can be bounded~(truncated) by intersection of the elliptical planes on the ground~\cite{UAV_geometric4,UAV_geometric7}. These topologies can help in deriving geometrical characteristics of the AG propagation scenario. The distribution of scattering objects, on land or water, can be modeled stochastically, and this concept can be used to create so-called geometrically-based stochastic channel models~(GBSCMs). For aircraft moving through an area above such a distribution, this gives rise to intermittent MPCs~\cite{UAV_meas9}, as also seen in vehicle-to-vehicle channels.

In order to describe the statistical characteristics of a fading channel, typically first and second order fading statistics are used. The majority of the AG literature discusses first order fading statistics. The second order statistics of envelope level crossing rate and average fade duration are discussed in \cite{UAV_meas1,UAV_geometric9}, but many authors address other second order properties, primarily correlation functions in the time or frequency domains.

In case of propagation over water the PL is similar to that of free space \cite{UAV_meas13}, with a strong surface reflection. The other MPCs from the water surface are weaker, and of approximately equal power and time-of-arrival (ToA), whereas MPCs from obstacles on the water surface, e.g., large ships, can be stronger.


\subsection{Antenna Configurations for UAV AG Propagation}
The antenna is one of the critical components for AG communications due to limited space, and limitations of the aerodynamic structure~\cite{Antenna_stat1,Antenna_stat2}. Factors that affect AG link performance are the number, type and orientation of the antennas used, as well as the UAV shape and material properties. 

The majority of AG channel measurements employ stand alone (single) antennas, whereas in~\cite{AG_meas2}, an antenna array is used. There are some SIMO and MIMO antenna configurations available in the literature for AG propagation measurements\cite{UAV_meas4,UAV_meas_mimo1}. Omni-directional antennas are most popular for vehicular communications due to their superior performance during motion, whereas directional antennas (having better range via directional gain) can perform poorly during motion due to mis-alignment losses. With high maneuverability of UAVs during flight, omni-directional antennas are generally better suited than directional antennas. A potential major drawback of any antenna on-board UAVs is the shadowing from the body of the UAV. Similarly, orientation of antennas on-board UAVs can affect the communication performance~\cite{UAV_802.11_1,UAV_802.11_2}.

The use of multiple antennas to enable diversity can yield spatial diversity gains even in sparse multipath environments~\cite{UAV_meas_mimo2,SIMO}. Similarly, multiple antennas can be used for spatial selectivity such as beam forming/steering. However, due to limited space on-board UAVs, space diversity using multiple antennas is difficult to achieve, especially for lower carrier frequencies. Beamforming using antenna arrays operating at mmWave frequencies, for example, can be used to overcome fading and improve coverage, but array processing will require high computational resources on-board. The employment of MIMO systems for enhancing the channel capacity of the AG propagation channel has been suggested in~\cite{UAV_geometric8,AG_geometric2}. By changing the diameter of a circular antenna array and the UAV flying altitude, different values of MIMO channel capacity were obtained~\cite{UAV_geometric8}. Whereas in~\cite{AG_geometric2}, optimizing the distance between the antenna elements using linear adaptive antenna arrays was proposed to increase MIMO channel capacity. %

\subsection{Doppler Effects}
Due to UAV motion, there are Doppler frequency shifts that depend on the velocity of the UAV and the geometry. Higher Doppler frequency presents a problem if the different signal paths are associated with largely different Doppler frequencies, yielding large Doppler spread. This can happen if the aircraft is relatively close to the GS.
If the aircraft is further away from the GS, and at sufficient altitude, the paths should all have a very similar Doppler frequency as the objects in the close surroundings of the GS causing MPCs are seen all under similar angles from the aircraft.
The effect of a large Doppler frequency that is constant for all MPCs should be well mitigated by  frequency synchronization. Doppler shifts can introduce carrier frequency offset~(CFO) and inter carrier interference, especially for orthogonal frequency division multiplexing~(OFDM) implementations. There are several studies that consider modeling of Doppler spread~\cite{UAV_doppler1,UAV_doppler2,AG_sim1,AG_meas7,manned_1,UAV_meas1,UAV_sim9,UAV_sim14}. Some channel access algorithms e.g., multi carrier code division multiple access, have been shown to be robust against Doppler spread in AG propagation~\cite{AG_meas4}.


\section{AG Channel Measurements: Configurations, Challenges, Scenarios, Waveforms} \label{AG_measurements}
Several AG channel measurement campaigns using piloted aircraft and UAVs have been recently reported in the literature. These measurements were conducted in different environments and with different measurement parameters. In this section, we provide a brief classification of these measurements based on environmental scenario, sounding signal, carrier frequency, bandwidth, and antenna specifications and placement. As available, we also provide UAV type and speed, heights of UAV and GS from terrain surface, link distance between transmitter~(TX) and receiver~(RX), elevation angle, and the channel statistics provided by the cited authors. These channel measurement parameters are given in Table~\ref{Table:Measurement}. 

\begin{table*}[htbp]
	\begin{center}
      \footnotesize
    	\caption{Important empirical AG channel measurement studies in the literature.} \label{Table:Measurement}
\hspace*{-.5cm}\begin{tabular}{@{} |P{.5cm}|P{1.4cm}|P{0.9cm}|P{0.9cm}|P{.6cm}|P{3.5cm}|P{.6cm}|P{1.5cm}|P{1.2cm}|P{0.8cm}|P{2.6cm}| @{}}
\hline
\textbf{Ref.}&\textbf{\hspace{0pt}Scenario}&\textbf{\hspace{0pt}Sound. Sig.}&\textbf{\hspace{0pt}Freq. (GHz)}& \textbf{\hspace{0pt}BW (MHz)}&\textbf{\hspace{0pt}Antenna and mounting}&\textbf{\hspace{0pt}\pmb{$P_{\rm T}$} (dBm)}&\textbf{\hspace{0pt}UAV, $\pmb{v_{\rm max}}$ (m/s)}&\textbf{\hspace{0pt}\pmb{$h_{U}$}, \pmb{$h_{G}$}, d (m)}&\textbf{\hspace{0pt}\pmb{$\theta$} (deg.)}&\textbf{\hspace{0pt}Channel statistics}\\
\hline
\cite{UAV_meas1}&{Urban} &\hspace{0pt}CW&$2$&$.0125$&$1$ Monopole on UAV for TX, $4$ on GS for RX&$27$&Air balloon, $8$&$170$, $1.5$, $6000$&$1$-$6$ &$P_{\rm R}$, Auto-correlation of direct and diffuse components \\
\hline
\cite{UAV_meas2}&{Open field, suburban}&PRN&$3.1$ - $5.3$&$2200$&$1$ Dipole on UAV for TX and $1$ on GS for RX &$14.5$&Quad-copter, $20$&$16$, $1.5$, $16.5$&-&PL, PDP, RMS-DS, TOA of MPCs, PSD of sub-bands \\
\hline
\cite{UAV_meas3,UAV_meas4,UAV_meas5,UAV_meas6,UAV_meas9,UAV_meas10,UAV_meas11,UAV_meas12,UAV_meas13,UAV_meas16}&{Urban, suburban, hilly, desert, fresh water, harbor, sea}&DS-SS&$0.968$, $5.06$&$5$, $50$&$1$ directional antenna on GS for TX, $4$ monopoles on UAV for RX&$40$&Fixed wing, $101$&$520-1952$, $20$, $1000-54390$&$1.5$-$48$&PL, PDP, RMS-DS, K-factor, tap probability and statistics (power, delay, duration) in TDL model\\
\hline
\cite{Schneckenburger_TAES2015}&rural, suburban&OFDM&$0.97$&$10$&$1$ monopole antenna on GS for TX, $1$ monopole on aircraft for RX&$37$&Fixed wing, $235$&$11000$, $23$, $350000$&$0$-$45$&PL, PDP, DPP\\
\hline
\cite{UAV_meas7}&{rural, suburban, urban, forest}&FMCW&$5.06$&$20$&$1$ monopole on UAV for TX, $1$ patch antenna on GS for RX&$30$&Fixed wing, $50$&-, $0$, $25000$&-&CIR, PG, RSS\\
\hline
\cite{UAV_meas8}&{Urban}&MSK&$2.3$&$6$&$1$ Whip antenna on UAV as transceiver, $1$ patch antennas as transceiver on GS&$33$&Fixed wing, $50$&$800$, $0.15$, $11000$&$4.15$-$86$&RSS\\
\hline	
\cite{UAV_meas15}&{Urban, suburban, rural}&GSM, UMTS&$0.9$, $1$, $9$-$2$, $2$&-&Transceiver on balloon and GS&$41.76$&Captive balloon&$450$, -, -&-&RSSI, handover analysis\\
\hline	
\cite{UAV_meas_mimo1}&{Urban, hilly, ocean}&OFDM&$2.4$&$4.375$&$4$ whip antennas on AV for TX, $4$ patch antennas on GS for RX&-&Fixed wing, $120$&$3500$, -, $50000$&-&Eigen values, beam-forming gain\\
\hline	
\cite{UAV_meas_mimo2}&{Rural}&PRN, BPSK&$0.915$&$10$&$2$ helical antennas on AV for TX, $8$ at GS for RX&$44.15$&Fixed wing, $36$&$200$, -, $870$&$13$-$80$&CIR, $P_{\rm R}$, RMS-DS, spatial diversity\\
\hline	
\cite{UAV_802.11_1}&{-}&OFDM&$5.28$&-&$4$ omni-directional on UAV for TX, $2$ on GS for RX&$18$&Fixed wing, $17.88$&$45.72$, $4.26$, -&-&$P_{\rm R}$, RSSI\\
\hline	
\cite{UAV_802.11_2}&{Urban, open field}&OFDM&$5.24$&-&$2$ omni-directional on UAV for TX, $2$ on GS for RX&$20$&Quad-copter,~$16$&$120$, $2$, $502.5$&-&RSSI\\
\hline	
\cite{UAV_802.11_3}&{Open field}&OFDM&$5.24$&-&$3$ omni-directional on UAV for TX, $3$ on GS for RX&$20$&Quad-copter,~$16$&$110$, $3$, $366.87$&$10$-$85$&RSS\\
\hline	
\cite{UAV_802.11_4}&{-}&IEEE $802.15.4$&$2.4$&-&On board inverted F transceiver antenna on UAV and GS&0&Hexacopter, $16$&$20$, $1.4$, $120$&-&RSSI\\
\hline	
\cite{UAV_challenges3}&{Suburban}&Wifi, $3$G/$4$G&-&-&Transceiver on UAV and GS&-&Hexacopter, $8$&$100$, -, -&-&$P_{\rm R}$, RTT of packets\\
\hline	
\cite{UAV_challenges4}&{ Forest (anechoic chamber)}&-&$8$-$18$&-&Spiral antennas on TX and RX&-&-&$2.3$, $0.6$, $2.85$&$26$-$45$&$P_{\rm R}$\\
\hline	
\cite{UAV_meas18}&{Open area}&Mod. sig.&$5.8$&-&$2$ Monopole, $1$ horn on UAV for TX, $2$ on GS for RX&-&Fixed wing,-&$150$, $0$, $500$&-&$P_{\rm R}$\\
\hline	
\cite{UAV_meas19}&{Open area/foliage}&$802.11$ b/g&$5.8$&-&$1$ omni-directional on GS for TX, $4$ on UAV for RX&-&Fixed wing,~$20$&$75$, $.2$, -&-&Diversity performance\\
\hline	
\cite{UAV_meas20}&{Urban/ suburban, open field, foliage}&CW&$2.00106$, $2.00086$&-&$2$ monopoles on UAV for TX, $2$ on GS for RX&$27$&Gondala airship,~$8.3$&$50$ and above, $1.5$, $2700$&$1$&$P_{\rm R}$\\
\hline	
\cite{UAV_meas22}&{Urban, rural, open field}&&$0.915$&-&$1$ omni-directional antenna on UAV for TX, $1$ on GS for RX&-&Quad-copter,-&-, $13.9$, $500$&-&RSSI, PL\\
\hline	
\cite{AG_meas1}&{Sea}&PRN&$5.7$&-&Omni-directional on AV for TX, $2$ directional antennas at GS for RX&$40$&Fixed wing AV,-&$1830$, $2.1$,$7.65$, $95000$&-&PL\\
\hline
\cite{AG_meas2}&{Urban}&CW&$2.05$&-&$1$ monopole on AV for TX, $4$ on GS for RX&-&Aerial platform,-&$975$, -, -&$7.5$-$30$&PDP, RMS-DS, MPCs count, $K$-factor, PL\\
\hline
\cite{AG_meas4}&{Near airport}&CW&$5.75$&-&Directional antenna on GS for TX and omni-directional on AV for RX&$33$&Fixed wing AV,-&$914$, $20$, $85000$&$80$&$P_{\rm R}$, Fading depth, $K$-factor, PL\\
\hline
\cite{AG_meas5}&{Urban, hilly}&Chirp&$5.12$&$20$&$1$ monopole antenna on GS for TX and $1$ omni-directional on AV for RX&$40$&Fixed wing AV,~$293$&$11000$, $18$, $142000$&$(-16)-5$&PDP\\
\hline

\end{tabular}
		\end{center}
		  \end{table*}

In the reported AG propagation measurements, either TX  or RX on UAV/GS is stationary. Measurements with both TX and RX moving for AG propagation are rare. A notable contribution of wide-band AG propagation measurements is available in the form of multiple campaigns conducted in the L and C bands using single-input-multiple-output~(SIMO) antenna configuration for different terrain types and over water/sea \cite{UAV_meas3,UAV_meas4,UAV_meas5,UAV_meas6,UAV_meas9,UAV_meas10,UAV_meas11,UAV_meas12,UAV_meas13,UAV_meas16}. The rest of the cited channel measurements are conducted in different frequency bands ranging from narrow-band to ultra-wideband~(UWB) with various types of sounding signals. 
\subsection{Channel Measurement Configurations}
These channel measurements used different types and configurations of antennas. The most commonly used antenna type is omni-directional and the most commonly used configuration is single-input-single-output (SISO). The positioning of an antenna on the UAV is important to avoid both shadowing from the airframe and  disruption of the aircraft's aerodynamics. In the majority of measurements the antennas were mounted on the bottom of the aircraft's fuselage or wings. The orientation of antennas on UAV and ground can also affect the signal characteristics~\cite{UAV_802.11_1,UAV_802.11_2,UAV_802.11_3,UAV_802.11_4}. This characteristic is most important during banking turns, and when the aircraft pitch angle deviates from horizontal. The elevation angle between TX and RX antennas is dependent on the height of UAV and GS and often continuously varies during the flight. 

In the majority of the communication applications envisioned for UAVs, the aerial node is expected to be stationary (or mostly so) in space for a given time. As noted, for communications with a mobile UAV, the velocity will affect the channel statistics. For UAVs operating at higher velocities, the coherence time of the channel decreases, and this translates into a larger Doppler spread. For connections to multiple UAVs, where hand-overs are required, this means that the number of handovers will also generally increase with velocity, and this will require additional processing. Additionally, higher velocities will result in increased air friction and mechanical turbulence that generally result in increased noise levels. Many of the AG channel measurements in the literature have been conducted with fixed wing aircraft with maximum speeds varying from $17$~m/s to $293$~m/s. The speed of rotorcraft and air balloons is much less than that of fixed wing aircraft, and ranges from 8~m/s to 20~m/s. 

The height of the UAV above ground is an important channel parameter and will also affect the channel characteristics. For example, increasing the height of the UAV usually results in reduced effect of MPCs~\cite{wahab_mmWave} from surrounding scatterers. Another benefit of higher UAV altitude is larger coverage area on the ground. Similarly, the height of GS will also affect the channel characteristics. For a given environmental scenario, there may be an optimal height of the GS~\cite{UAV_meas2},e g., this might be a balancing of attenuation and multipath diversity.

Example propagation measurements using rotorcraft and air balloons during flight and hovering are available in~\cite{UAV_meas1,UAV_meas2,UAV_802.11_2,UAV_802.11_4}. These AG propagation measurements were obtained at different UAV heights ranging from~$16$~m to $11$~km, and link distances~$16.5$~m to $142$~km. The UAV latitude, longitude, yaw, pitch, and roll readings are typically obtained from GPS RXs and often stored on-board. 

Apart from conventional AG channel sounding, there are some indirect UAV AG channel measurements available from use of radios employing different versions of protocols of the IEEE~802.11 standards~\cite{UAV_802.11_1,UAV_802.11_2,UAV_802.11_3,UAV_802.11_4}. The IEEE $802.11$ supported devices offer a very flexible platform and may provide insight for UAV deployments in different topologies and applications, e.g., UAV swarms. Yet because of the specific features of $802.11$, the resulting measurements are applicable to particular protocol setup and radio configuration, and rarely provide detailed propagation channel characteristics. 

Air-to-air (AA) communications with UAVs has not been studied extensively in the literature~\cite{air_air}. The AA communications is particularly important for scenarios where multiple drones communicate among a swarm. This swarm then usually communicates with one or more GS via a back-haul link from one or several of the UAVs. The AA communications is similar to free space with a strong LOS and often a weak ground reflection, but this is dependent on the flight altitude and environment. The communication channel is mostly non-dispersive for higher altitudes but can be rapidly time-varying, dependent on the relative velocities of the UAVs and the scattering environment ~\cite{UAV_air_to_air}.

 \subsection{Challenges in AG Channel Measurements}
There are many challenges in AG channel measurement campaigns as compared to terrestrial measurements. The biggest challenges are the payload limitation of the UAVs, and the operating range and height of UAVs, which in the USA is set by the FAA~\cite{FAA}. Larger UAVs also incur larger test costs. Due to restrictions on the height of UAVs above ground, UAVs at lower altitudes have lower LOS probability and are hence more susceptible to shadowing, especially in suburban and urban areas. Due to limitations on payload, higher transmit power measurements on-board the UAVs are difficult to achieve, and similarly, complex RX processing on-board UAVs can consume a prohibitive amount of power. 

\begin{table*}[htbp]
	\begin{center}
      \footnotesize
		\caption{UAV AG propagation characteristics for five different scenarios.} \label{Table:UAV_environment_scenarios}
\label{Table:Table_III}
\hspace*{-.5cm}\begin{tabular}{@{}|P{2.2cm}|P{7.0cm}|P{7.0cm}|@{}}
 \hline
\textbf{Scenario}& \textbf{Characteristics of scenario}& \textbf{Important factors} \\
\hline
{Urban/suburban} &Ratio of land area vs ratio of open to built-up area, distribution of building sizes and heights, distribution of ground terminals (vehicles, pedestrians), distribution and characteristics of vegetation, water bodies, etc. & Material of buildings and rooftops\\
\hline
{Rural/open field}&Type and density of vegetation, distribution and sizes of the sparse buildings &Surface roughness, soil type, and moisture content\\
\hline
{Hilly/mountainous}&Terrain heights and slopes, distribution and type of vegetation, distribution and sizes of buildings&Ground slope, ground roughness \\
\hline
{Forest}&Density and types of foliage, and height distributions&Leaves and branches distribution \\
\hline
{Over water}&Water type (sea or fresh), distributions and sizes of water surface objects (boats, platforms, etc.), distributions of littoral objects (buildings, water tanks, etc.), and water surface variation (e.g., sea state)& Modified reflection coefficient as compared to ground, ducting effect in case of over sea \\
\hline

\end{tabular}
		\end{center}
			\end{table*}

Other challenges include varying conditions of the terrain during flight, meteorological conditions~(winds and rain), antenna positioning on the UAV, precise location measurement of UAVs in space over time, diverse telemetry control for different types of UAVs having specific latencies, bandwidth and reliability issues, and limited flight time for most small UAVs due to limited battery life~\cite{UAV_challenges1, UAV_challenges2, UAV_challenges3}. Due to the motion of UAVs in three dimensional space, it is challenging to precisely measure the distance between the UAV and the GS. Momentary wind gusts that cause sudden shifts in UAV position can make it difficult to accurately track the UAV path. The most common technique of measuring the instantaneous distance is by using global positioning system (GPS) traces on both the UAV and GS, but of course GPS devices have accuracy limitations and navigation signals may also be susceptible to interference in different flying zones. 

\subsection{AG Propagation Scenarios}
A typical type of terrestrial channel sounding equipment, a vector network analyzer, cannot be used for UAV based AG channel sounding due to payload constraints, physical synchronization link requirements, and UAV mobility~\cite{VNA}. Therefore, channel sounding for both narrow-band and wide-band channels using impulse, correlative, or chirp sounding techniques are employed, where the RX is typically on the ground due to payload and processing constraints. 

\begin{figure*}[!h]
	\centering
	\includegraphics[width=1.5\columnwidth]{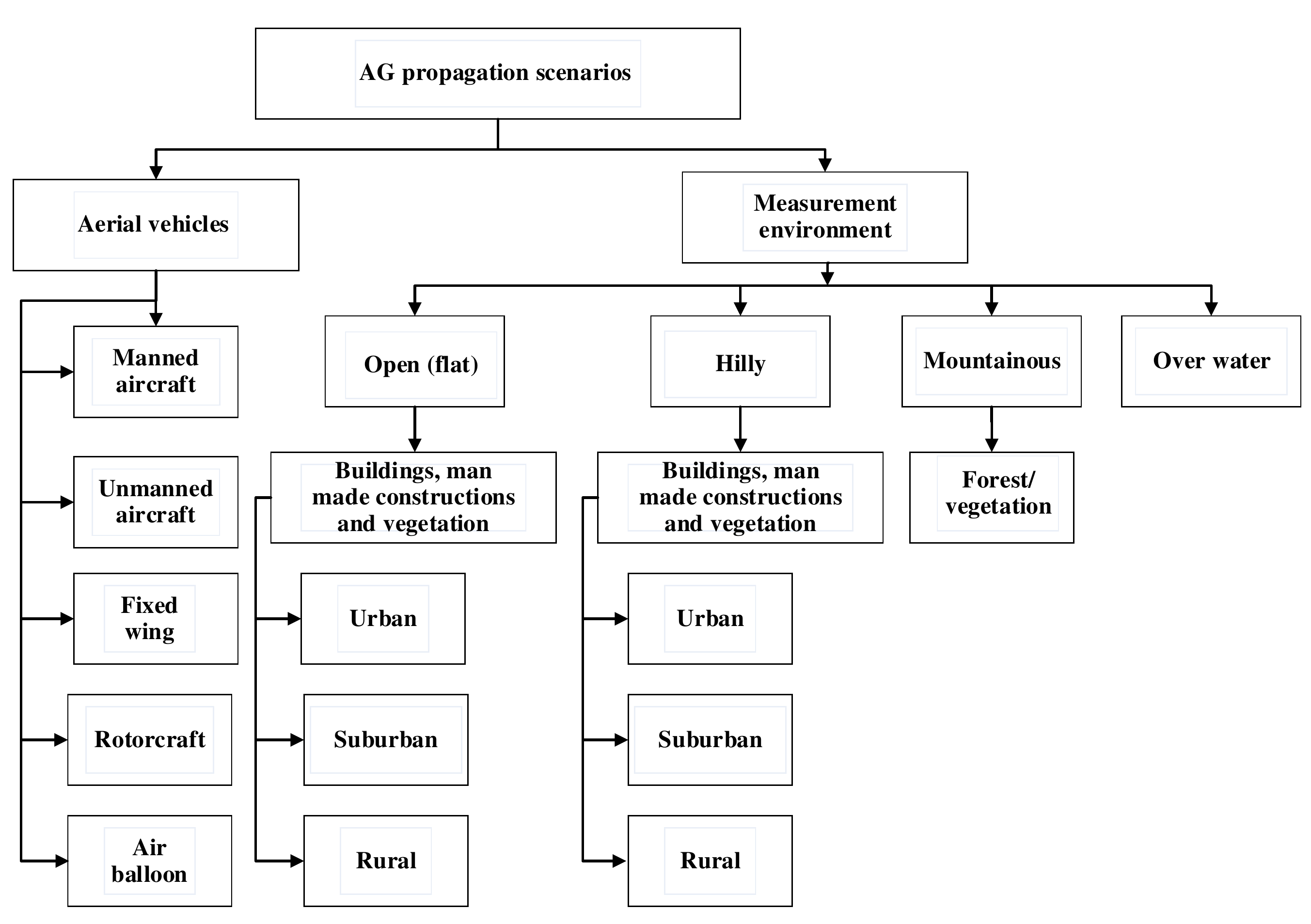}
	\caption{Measurement scenarios for UAV AG propagation channel.}\label{Fig:scenarios}
\end{figure*}

Proper selection of channel measurement parameters in a given environment is critical for obtaining accurate channel statistics for a given application. The AG propagation environment is generally classifed on the basis of the terrain type, namely flat, hilly, mountainous, and over water. A particular terrain can have a given cover e.g. grass, forest, or buildings. The most widely accepted terrain cover classification is provided by the International Telecommunication Union~(ITU)~\cite{Cover_terrain}. In this survey we classify the cited measurement scenarios as open (flat), hilly/mountainous, and over water. Each scenario can be subdivided on the basis of the terrain cover as shown in Fig.~\ref{Fig:scenarios}.

For any environment, different types of radio controlled UAVs can be used. Balloons or dirigibles are simple to operate but do not have robust movement characteristics. The non-balloon UAVs can be broadly classified as fixed wing and rotorcraft. The fixed wing UAVs can glide and attain higher air speeds and generally travel farther than the rotorcraft, but rotorcraft are more agile, e.g., most can move straight vertically. Rotorcraft also have the ability to hover, which is not possible for nearly all fixed wing UAVs. The UAV AG propagation scenarios in different environments with particular characteristics are described in Table~\ref{Table:UAV_environment_scenarios}. In the rest of this subsection, we review the different AG measurement scenarios depicted in Fig.~\ref{Fig:scenarios}.

\begin{figure*}
\includegraphics[width=\textwidth]{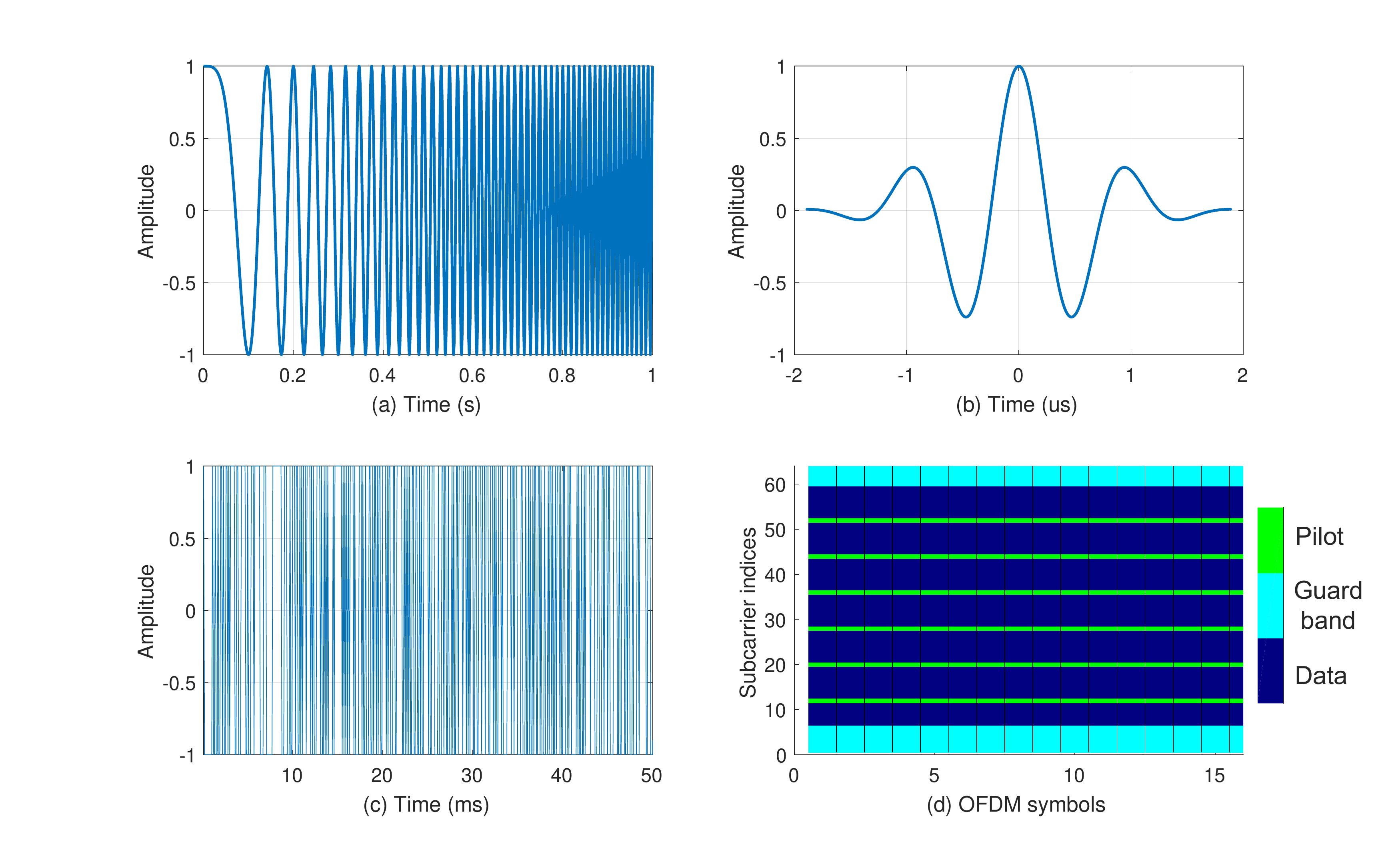}
\caption{Sounding signals (a) Chirp signal, (b) Short duration Gaussian pulse sounding signal at center frequency of $1$~MHz and fractional bandwidth of $60$$\%$, (c) PRN sequence of polynomial degree 10 shown half of the overall period, (d) OFDM sounding signal resource mapping with $64$ sub-carriers, $16$ symbols and $6$ pilots.}
\label{Fig:Sounding_signals}
\end{figure*}

\subsubsection{Open Space}
A major part of the literature on AG propagation covers open (flat) terrain. This open terrain can have different terrain covers that affect the channel characteristics. One of the major terrain cover types is  buildings. The distribution of building sizes, heights, and their areawise densities allows sub-classification into urban, suburban and rural areas as depicted in Fig~\ref{Fig:scenarios}. In case of urban and suburban areas, there is a higher concentration of man made structures in a given space, e.g., buildings, roads, bridges, large signs, etc. The distribution (and composition) of these complex scatterer structures can strongly influence the channel characteristics. In rural areas, typically buildings are sparse, and of lower height than in urban settings, although large warehouses and other structures could yield strong MPCs. 


\subsubsection{Hilly/Mountainous}
The hilly/mountainous terrain is characterized by uneven ground heights; equivalently, a large standard deviation of terrain height. The propagation PL in hilly and mountainous areas will mostly follow the two ray model with adjustments due to surface roughness, and potentially reflections from smooth sections of mountain slopes or an occasional large building. The PL over or beyond terrain obstructions can employ established models for diffraction, e.g.,~\cite{ITU_mountain} but with first Fresnel zone clearance between TX and RX, PL is close to free space~\cite{UAV_meas10,UAV_meas6}. Channel dispersion, typically quantified by the RMS-DS, is generally smaller than in urban/suburban environments~\cite{UAV_meas10} but can be large if a strong reflection occurs from a large and distant mountain slope. Generally, hilly and mountainous settings present fewer reflections than more populated regions because of the absence of large numbers of nearby scatterers.  

\subsubsection{Forest}
There are few comprehensive studies covering AG propagation in forests, especially with UAVs, although there are numerous publications for \textit{roadside} shadowing for satellite channels, e.g.,~\cite{forest_satellite1, forest_satellite2, forest_satellite3}. In these studies, propagation effects--typically attenuation--from particular volumes of trees, along with temporal fade statistics are analyzed for long range AG communications. Generally for AG propagation with a GS within a forest, the channel characteristics are dominated by the type and density of trees. Small UAVs within a forest experience different scattering characteristics depending upon height, e.g., the scattering near the tree trunk will be different from that near the tree crown \cite{UAV_challenges4}. The scattering is also dependent on the type and density of leaves and branches of the trees, and hence for deciduous trees, can vary seasonally.  

\subsubsection{Water/Sea}
The AG propagation channel for over water settings is similar to that for open settings, with different surface reflectivity and roughness than ground. The PL can be represented using a two ray PL model, with variations attributable to surface roughness (see small-scale fading in the following section). The RMS-DS in this case is generally smaller than in environments with a large number of obstacles (urban, suburban), although if large objects are on or just off shore, these may produce significant reflections and large delay spreads if geometry permits. 

In case of propagation over sea, the height of waves in a rough sea can introduce additional scattering and even diffraction for very low height stations on the sea. An interesting propagation phenomenon that can also occur over sea is ducting, where anomalous index of refraction variation with height results in propagation loss less than that of free space~\cite{AG_meas1}. This phenomenon is dependent on frequency and meteorological conditions, and is thus typically addressed statistically~\cite{Duct_ITU}.

\subsection{AG Channel Sounding Waveforms}
As noted in~\cite{AG_meas5,AG_meas7}, common channel sounding signals include short pulses~(approximately impulses), direct sequence spread spectrum~(DS-SS) signals for correlative processing, linearly varying frequency modulation~(chirp) signals, and multi-tone signals. Different example sounding signals are shown in Fig.~\ref{Fig:Sounding_signals} representing a chirp signal, RF Gaussian pulse, pseudo-random number~(PRN) sequence, and orthogonal frequency division multiplexing~(OFDM) sounding signals. These sounding signals have been used in different measurement campaigns summarized in Table~\ref{Table:Measurement} in different AG channel measurement scenarios given in Fig.~\ref{Fig:scenarios}. Short duration pulses are direct approximations of input impulses and MPCs can be directly measured in the time domain (e.g., via a sampling oscilloscope). The primary drawback is generation of sufficient pulse energies to reach long distances, and large peak-to-average power ratios (PAPR). The DS-SS signaling uses pseudo-random (PR) sequences to generate a wideband noise-like signal that is demodulated with a sliding~(or sometimes a stepped) correlator; this correlation processing yields an estimate of the channel impulse response (CIR). The DS-SS technique can use binary phase shift keying transmission and with modest filtering this yields a low PAPR. Chirp sounding has the advantage of high frequency resolution and the potential to sweep over large frequency ranges; PAPR can be the ideal value of unity. The chirp technique yields the channel transfer function, from which the CIR is obtained via inverse Fourier transformation.

Another popular technique is the use of a multitone signal, with the idea of sampling the channel transfer function. This is in essence an OFDM based channel sounding. One advantage of using OFDM sounding is that known data can be used for sounding, hence allowing some data transmission along with channel sounding \cite{ofdm_sounding}. The OFDM signals have the advantage of a flat spectrum but of course a \textit{sinc} ($\sin(x)/x$) delay domain response and a large PAPR. Details on these various sounding signals can be found in the literature, e.g., \cite{parsons}.

Different carrier frequencies can be used to sound the AG channel and in principle this is completely arbitrary, but most measurements aim at frequency bands in which UAV use is at least possible. Measurements have ranged from $100$~MHz to $18$~GHz with perhaps most of the measurements carried out in the $5$~GHz band ($5.06$~GHz - $5.8$~GHz). Similarly, sounding signal bandwidth varies, from very narrow-band to several tens of MHz or more. In~\cite{UAV_meas2}, UWB channel sounding with a bandwidth of $2.2$~GHz was used, yielding sub-nanosecond time resolution.



\section{UAV AG Propagation Measurement/Simulation Results in the Literature}

Several types of channel statistics are useful for characterizing the channel for different applications. For AG propagation, the channel statistics are similar to those gathered for terrestrial channels. In general, propagation channels are linear and time varying, but can sometimes be approximated or modeled as time-invariant. For linearly time-varying channels, the CIR or its Fourier transform, the time varying channel transfer function (CTF), completely characterizes the channel~\cite{UAV_meas1, UAV_meas3,UAV_meas4,UAV_meas5,UAV_meas6,UAV_meas9,UAV_meas10,UAV_meas11,UAV_meas12,UAV_meas13,UAV_meas16,UAV_meas_mimo2, AG_meas2, AG_meas5}. As noted, due to relative motion of the UAV, the AG channel may be stationary only for small distances~\cite{UAV_meas4}. Thus, \textit{stationary distance} needs to be taken into account when estimating the channel statistics~\cite{WSS_dist1, WSS_dist2,UAV_meas13}. 

Another higher-level parameter that has been used by some researchers to characterize the quality of the AG propagation channel is throughput, but of course this is highly dependent upon the transmitter and receiver implementation, and parameters of the air interface, such as the number of antennas and the transmit power. Hence this measure is of limited use for assessing the AG channel itself. Similarly, for MIMO channels, beam-forming gain, diversity, and capacity of the channel are often estimated. Some commonly reported channel characteristics for AG propagation channels are given in the following subsections.

\subsection{Path Loss/Shadowing}
Most of the AG propagation campaigns address PL and if present, shadowing, in different scenarios. For AG channels with an LOS component, PL  modeling begins with free space propagation loss; when the earth surface reflection is present (not blocked or suppressed via directional antennas), path loss can be described by the well-known two-ray model. Parallel to the developments in terrestrial settings, most of the measurements employ the log-distance PL model where the loss increase with distance is indicated by the path loss exponent (PLE). In~\cite{UAV_meas2}, PL is calculated for open field and suburban areas for different UAV and GS heights for a small hovering UAV. Comprehensive PL measurements in L and C bands were carried out in different propagation scenarios in~\cite{UAV_meas3,UAV_meas4,UAV_meas5,UAV_meas6,UAV_meas9,UAV_meas10,UAV_meas11,UAV_meas12,UAV_meas13,UAV_meas16} as summarized in Table~\ref{Table:Measurement}. The values of PLE were found to be slightly different for urban, suburban, hilly, and over water scenarios, but are generally close to the free-space value of 2 with standard deviation around the linear fit typically less than 3 dB. 

In~\cite{UAV_802.11_2}, it was observed that the PLEs for IEEE 802.11 communications were different during UAV hovering and moving due to different orientations of the on-board UAV antennas. Therefore, antenna patterns can distort the true channel PL characteristics and removing their effect is not always easy or possible. On the other hand, for the specific UAV configuration used, the resulting PL model is still useful. Typically, PL for LOS and NLOS conditions are provided separately, e.g.~\cite{UAV_geometric1}, where for the NLOS case, there is an additional small-scale (often modeled as Rayleigh) fading term, and a constant reflection term in addition to the LOS PL. Analogously, the LOS models for L- and C-bands can incorporate Ricean small scale effects~\cite{UAV_meas4}. In~\cite{UAV_sim2}, the reported PL is described as a function of the elevation angle between the low altitude platform and RX $\theta_{\rm e}$ given as follows:
\begin{equation}
FSPL = 20\log\bigg(\frac{\Delta h}{\sin\theta_{\rm e}}\bigg) + 20\log(f_{\rm MHz}) - 27.55,
\end{equation}
where $\Delta h = h_{\rm LP} - h_{\rm Rx}$ is the difference between the height of the low altitude platform and the RX on the ground. The argument $\Delta h/\sin\theta_e$ is simply the link distance expressed as a function of elevation angle.

Path loss including shadowing is reported in~\cite{UAV_meas1, UAV_meas14, AG_meas2, AG_meas4, UAV_meas2, UAV_sim4}, where we note that in LOS cases without actual obstruction of the first Fresnel zone, the physical mechanism causing PL variation is not actually shadowing but often small-scale effects. In \cite{UAV_meas1}, PL and its associated shadowing was attributed to buildings only when the UAV was flying near the ground whereas when flying higher, actual shadowing was not present but variation from small-scale fading still occurred. One can also estimate losses due to ``partial'' shadowing by conventional methods. For example the shadowing in~\cite{UAV_meas14} was found to be a function of elevation angle, where the shadowing magnitude was estimated by using the uniform theory of diffraction. 

\begin{figure}
\includegraphics[width=0.9\columnwidth]{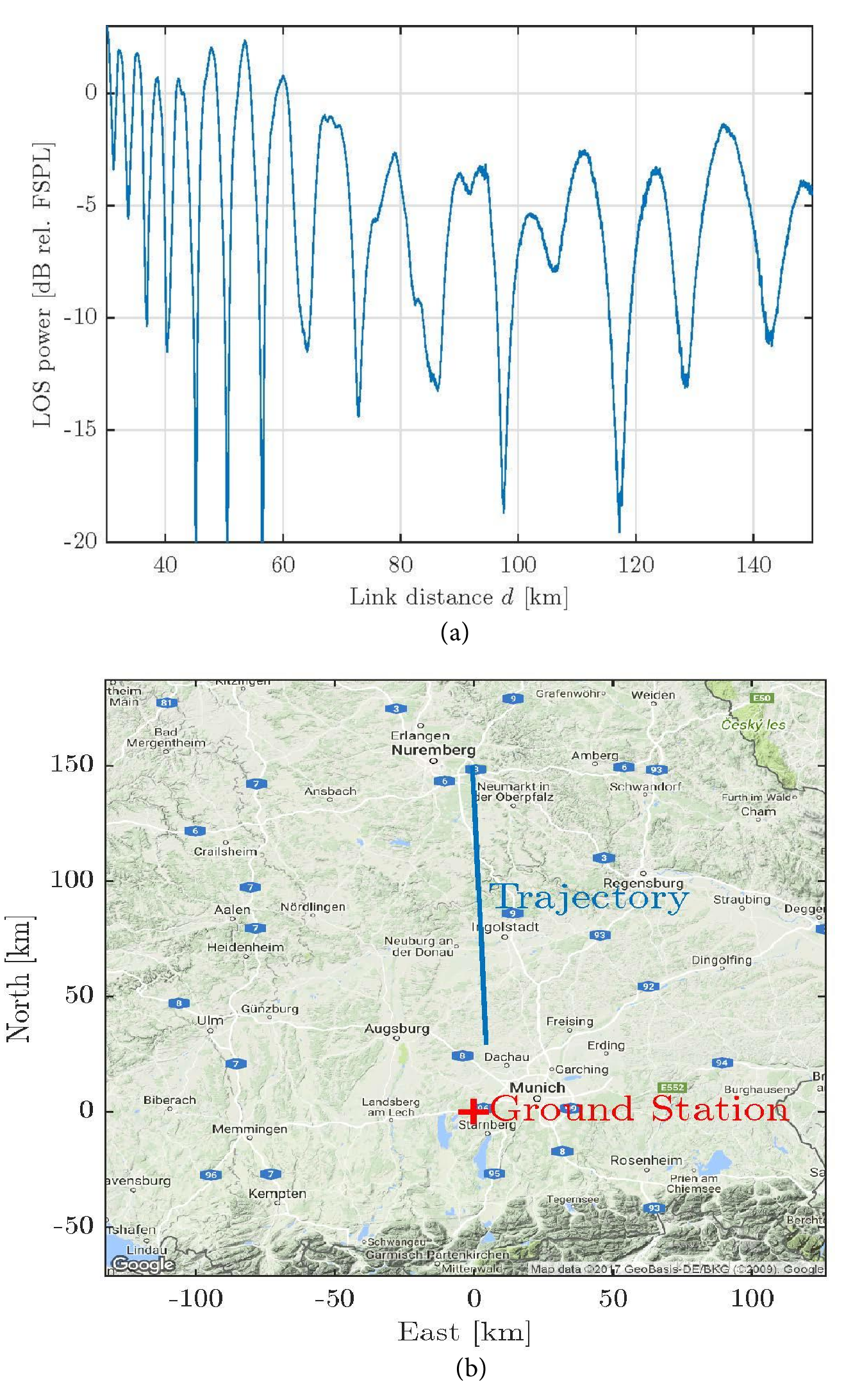}
\caption{(a) The LOS signal power variation due to ground multipath propagation. The power is normalized to free space path loss, (b) Measurement scenario environment in \cite{Schneckenburger_Eucap16_GroundMp}.}
\label{fig:GMP}
\end{figure}

In Fig.~\ref{fig:GMP}(a) we show an example for the variation of the LOS signal power due to ground reflected MPCs versus the link distance $d$. Specifically, this is the $combined$ effect of the LOS component and the unresolved ground reflection.
The measurements were taken in a rural environment using a 10 MHz signal bandwidth. The GS height $h_{\rm G}$ is $23$~m. The UAV trajectory is shown in Fig.~\ref{fig:GMP}(b). During the measurements, the specular reflection point first passed over the roof of a building and then over open grassy fields \cite{Schneckenburger_Eucap16_GroundMp}. From Fig.~\ref{fig:GMP}(a) we observe a periodic variation of the received power: an attenuation of the signal by more than $10$~dB is not uncommon. These signal fades will of coursed generally negatively impact the performance of any communication system. For an increasing link distance the frequency of the variation decreases--a direct manifestation of the two-ray model. Thus in such a channel, even for a UAV flying at a high speed a fade can easily last several seconds. It is essential to note that a ground MPC may not always be present, e.g. for the case when the ground is a poorly reflecting ground surface, or the surface is very rough relative to the signal wavelength.

The PL provides complete information on link attenuation, but another indirect parameter often used for channel attenuation estimation is received signal strength~(RSS). In~\cite{UAV_802.11_1,UAV_802.11_2,UAV_802.11_3}, RSS indicator data for an AG propagation channel based on IEEE 802.11a transmissions with different antenna orientations was provided. Data on fluctuations in  RSS due to multipath fading from tall building reflections was provided in~\cite{UAV_meas7}, where the RSS was found to decrease due to polarization mismatch between the TX and RX antennas when the aerial vehicle made a banking turn. The accuracy of RSS values in commercial products can vary considerably, so when these are used, care should be taken in calibration. 

\subsection{Delay Dispersion}
The power delay profile (PDP) is the "power version" of the CIR. This can be computed "instantaneously," or more traditionally, as an average over a given spatial volume (where the channel can be considered WSS). Various AG propagation studies in different environments have measured PDPs, and via the PDP the most common estimate of the delay-domain dispersion is estimated: the RMS-DS. Other dispersion measures such as the delay window or delay interval are also sometimes reported. Statistics for the RMS-DS statistic itself are often computed, e.g., in~\cite{AG_meas2}, mean RMS-DS values for different elevation angles was reported. As generally expected from geometry, the RMS-DS was found to decrease as elevation angle increases. In~\cite{UAV_meas2} PDPs were measured for open areas, suburban areas, and areas covered with foliage. 

The Saleh-Valenzuela model, originally developed for indoor channels, is sometimes used to model the PDP when MPCs appear grouped or "clustered" in delay. This model specifies the MPCs by such clusters, and the number of clusters is different for different environmental scenarios. PDPs were measured for different environments in~\cite{UAV_meas5,UAV_meas6,UAV_meas11,UAV_meas12,UAV_meas13,UAV_meas16,UAV_meas10}, and resulting RMS-DS statistics were provided. As expected, the delay spread was found to be dependent on the terrain cover with maximum delay spread values of $4\mu s$ for urban and suburban settings. The largest RMS-DS values generally occur when there are large buildings that can provide strong MPC reflections. For hilly and mountainous terrain,  maximum RMS-DS values of $1\mu s$ for hilly regions and $180 ns$ for the mountainous terrain were reported.  In over water settings, the maximum RMS-DS value reported was $350 ns$. Again, in all these settings cited here, a LOS component was present between GS and UAV, hence for the majority of the time, RMS-DS was small, on the order of a few tens of nanoseconds. In~\cite{UAV_sim8}, a finite-difference time domain model for the electric field propagating at very low heights over sea was developed. An RMS delay spread model for very high frequency (VHF) to $3$~GHz was presented, with RMS-DS a function of wave height. 

\subsection{Narrowband Fading Severity: Ricean $K$-factor}
Small scale amplitude fading in AG propagation channels usually follows a Ricean distribution due to the presence of a LOS component. The Ricean $K$-factor is defined as the ratio of dominant channel component power to the power in the sum of all other received components. The K-factor is often used to characterize the AG channel amplitude fading. In~\cite{AG_meas2}, as generally expected, the authors found that the $K$-factor increased with increasing elevation angle. The Ricean $K$-factor as a function of link distance was given in~\cite{AG_meas4}, during multiple phases of flight (parking and taxiing, take off and landing, and en-route). The en-route phase showed the largest $K$-factor, followed by take off and landing, and parking and taxiing. In~\cite{UAV_challenges4}, it was observed that the $K$-factor will differ with different types of scattering trees: values of $K$ ranging from $2$~dB to $10$~dB were reported. 

The $K$-factor was measured for both L-band and C-band AG propagation in~\cite{UAV_meas4, UAV_meas9, UAV_meas10, UAV_meas16} for urban, suburban, hilly and mountainous settings, and also for over fresh water and sea scenarios. The mean values of $K$-factor for urban areas were reported to be 12~dB and 27.4~dB for L-band and C-band respectively. The mean K-factor values for hilly and mountainous terrain was reported to be 12.8~dB and 29.4~dB for L-band and C-band respectively, whereas for over sea settings, $K$-factor mean values for L-band and C-band were found to be 12.5~dB and 31.3~dB, respectively. Worth pointing out is that in these ``strong LOS'' channels, the $K$-factor does not strongly depend on the GS environment. Also observed was that the C-band $K$-factor was larger than the L-band K-factor in all environments. This is attributable to two causes: first, the C-band measurement signal bandwidth was larger than that of L-band, ameliorating fading, and second, for any given incident angle and surface roughness (e.g., ground, or ocean), as carrier frequency increases, the surface roughness with respect to the wavelength also increases, and hence incident signals are scattered in multiple directions rather than being reflected in a single direction (toward the receiver). With fewer and/or weaker MPCs at the higher frequency, the $K$-factor is larger.

\subsection{Doppler Spread}
The Doppler effect is a well-known phenomenon for wireless mobile communications. Considering AG propagation with UAVs in a multipath environment, if we let $\phi_{i}$ represent the angle between the aircraft velocity vector and the direction from which the $i^{\rm th}$ MPC is received, the Doppler frequency shift of this $i^{\rm th}$ MPC is $f_{\rm d}^{i} = \frac{v\cos\phi_{i}}{\lambda}$, where $v$ is the UAV velocity, and $\lambda$ is the wavelength of the radio wave. (We assume here that the GS is motionless, else a more general formulation for the Doppler shift must be used.) If MPCs are received with different Doppler frequencies this phenomenon produces spectral broadening, called Doppler spread. 

In~\cite{AG_meas4,manned_1}, simulations were used to find the Doppler shift and its effect on the channel at different phases of flight~(parking and taxiing, en-route, and take off and landing). Doppler spread in a multipath environment implementing OFDM systems was considered in~\cite{AG_meas7}, where  arriving MPCs were observed to have different frequency offsets. In such a case, if the receiver CFO synchronizer cannot mitigate the effect of these different frequency offsets, this results in inter-carrier interference~(ICI). In~\cite{UAV_doppler2}, a mitigation technique for Doppler shift was proposed for the case where the UAV is relaying between two communication nodes. The UAV acts as a repeater that provides the required frequency shift to mitigate the Doppler effect. A three dimensional AG Doppler delay spread model was provided in \cite{UAV_doppler1} for high scattering scenarios. Doppler spread for AG propagation is also discussed in~\cite{AG_meas5, AG_sim1,UAV_meas_mimo2,UAV_meas2, UAV_meas3}.

\subsection{Measured Air Interface Statistics}
Apart from the main channel characteristics, there are other performance indicators that can be measured. Two of these are throughput and bit error ratio (BER) with particular communication technologies. As with RSSI measurements, these are useful for the particular technology and environment in question, but may offer very little that is directly relevant to modeling the AG channel. The throughput of an AG propagation channel was investigated in several studies, most commonly using the IEEE 802.11 protocol. Throughput analysis using different versions of the IEEE 802.11 protocol were carried out in~\cite{UAV_802.11_1,UAV_802.11_2}, for different antenna orientations, propagation distances, and UAV elevations. A throughput analysis of  IEEE~$802.11$n was carried out in~\cite{UAV_challenges1}, where--as expected--it was found that throughput is directly dependent on the modulation and coding scheme. Throughput analysis for data relaying and ferrying for an AG propagation channel was carried out in~\cite{UAV_challenges2}. It was observed here that mobile relaying can achieve more than twice the throughput of static relaying for a given delay tolerant system. 

Some results for BER as a function of signal-to-noise-ratio (SNR) for AG propagation channels are available in the literature to compare the performance of different implementation schemes. In~\cite{UAV_sim12}, BER was measured against SNR for different modes of LDACS1, as a function of distance and for different phases of flight. A similar study was conducted in~\cite{UAV_sim13}, where BER was measured against SNR for an over sea AG propagation channel with distance measuring equipment (DME) co-channel interference present. In~\cite{UAV_geometric6}, BER versus SNR analysis was performed for different flight route phases for different values of Ricean $K$-factor. BER versus SNR analysis was performed in~\cite{UAV_sim14} for comparing the effect of presence and absence of ICI for an IEEE~$802.11$a OFDM system in the presence of additive white Gaussian noise (AWGN).

 \subsection{Simulations for Channel Characterization}
Apart from measurement campaigns for AG propagation channel modeling, some simulation based channel characterizations are also available in the literature, where the real time environmental scenarios are imitated using computer simulations. Simulations in urban/suburban areas were performed in \cite{UAV_sim2, UAV_sim4,UAV_geometric6, UAV_sim9}. The antenna considered in these environments was omni-directional. Different carrier frequencies~$200$~MHz, $700$~MHz, $1$~GHz, $2$~GHz, $2.5$~GHz, $5$~GHz, and $5.8$~GHz were covered for AG channel characterization in the urban/suburban environments, and different heights of UAVs, ranging from $200$~m to 2000~m were considered. The PL (from simulated RSS) was estimated. Over sea based channel simulations were carried out in~\cite{UAV_sim8}, where a channel simulator imitating the sea environment was developed. Carrier frequencies from $3$~kHz-$3$~GHz were used, with the TX and RX placed $3.75$~m above the sea surface. The main goal of the study was to quantify sea surface shadowing for the marine communication channel using UAVs. The channel characteristics of PL and root mean square-delay spread (RMS-DS) were modeled based on the sea surface height. 

In~\cite{UAV_sim13}, simulations were conducted in environmental scenarios consisting of over sea, hilly, and mountainous terrain. Performance of AG communications using filter bank multicarrier~(FMBC) modulation systems and LDACS were compared. The results showed that FMBC has better performance than LDACs, especially in the presence of interference from DME signals. In the presence of the AG channel, the FBMC and LDACS performance is comparable. Other simulations of communication systems employed over AG propagation channels, for particular simulation scenarios, are also available in the literature~\cite{UAV_geometric5,UAV_doppler2,AG_meas7}.

In \cite{Optimum_height_UAV}, the effect of the UAV height for optimal coverage radius was considered. It is observed that by adjusting UAV altitude, outage probability can be minimized: a larger "footprint" is produced with a higher UAV altitude, but of course increased altitude can increase PL. An optimum UAV height is evaluated that maximizes the coverage area for a given SNR threshold. The Ricean $K$-factor was found to increase exponentially with elevation angle between UAV and GS, given as $K = c_1\exp^{c_2\theta}$, where $c_1$ and $c_2$ are constants dependent on the environment and system parameters. 
The relation between minimizing outage probability or maximizing coverage area for a given SNR threshold is solved only based on path loss without considering the effect of scatterers in the environment. The consideration of geometry of scatterers in the analysis would of course make it more robust and realistic.


\section{UAV AG Propagation Models}
\begin{figure*}[!t]
	\centering
	\includegraphics[width=1.5\columnwidth]{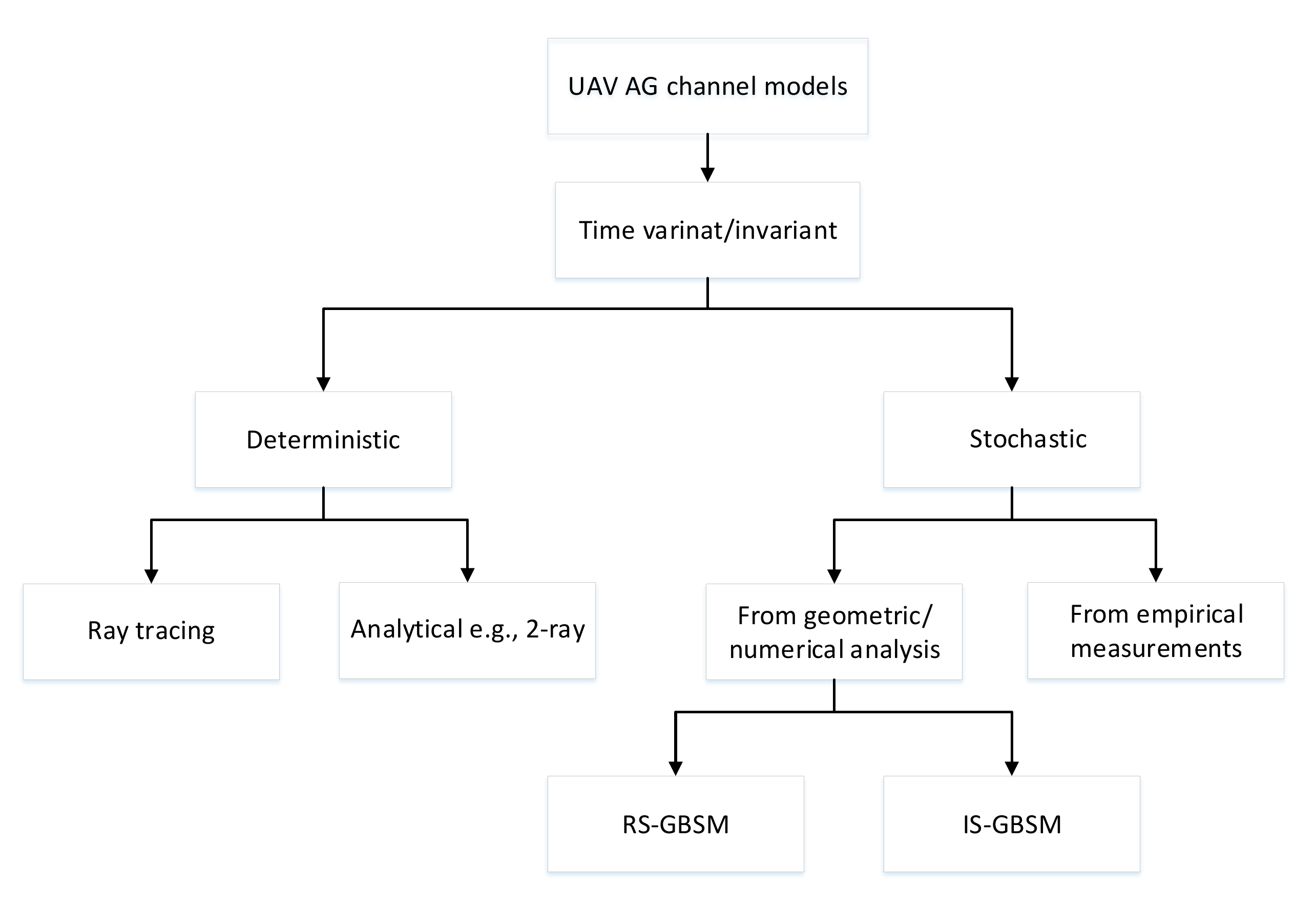}
	\caption{UAV AG channel model characterization.}\label{Fig:Channel_model}
\end{figure*}

The UAV AG propagation measurements discussed in the previous section are useful for developing models for different environments. In the literature, UAV AG propagation channel models have been developed using deterministic or statistical approaches, or their combination. These channel models can be for narrow-band, wide-band, or even UWB communications. Complete channel models include both large scale and small scale effects. In this section, we categorize AG propagation channel models in the literature as shown in Fig.~\ref{Fig:Channel_model}, and review some of the important channel models.

\subsection{AG propagation Channel Model Types}\label{Subsection:Channel_models}
Time-variant channel models can be obtained via deterministic or stochastic methods or by their combination. The deterministic methods often use ray tracing (or, geometry) to estimate the CIR in a given environment. These deterministic channel models can have very high accuracy but require extensive data to characterize the real environment. This includes the sizes, shapes, and locations of all obstacles in the environment, along with the electrical properties (permittivity, conductivity) of all materials. Hence such models are inherently site-specific. They also tend to require adjustment of parameters when comparing with measurement data. Since ray tracing based techniques employ high-frequency approximations, they are not always accurate. They are not as accurate as full wave electromagnetic solutions, e.g., the method of moments and finite difference time domain methods for solving Maxwell equations~\cite{raytracing}, but ray tracing methods are of course far less complex than these full-wave solutions. Such deterministic simulators are also very complex when they are used to model time varying channels. Ray tracing was used in~\cite{UAV_doppler1, UAV_meas15, UAV_sim2, UAV_sim4, UAV_sim16, UAV_sim18} for different fully deterministic AG propagation scenarios. 

The models in~\cite{UAV_meas10,UAV_meas12,UAV_meas13} are a mix of deterministic and stochastic models (sometimes termed quasi-deterministic). Specifically, the LOS and earth surface reflection are modeled deterministically via geometry, and the remaining MPCs are modeled stochastically, with parameter distributions (for MPC amplitude, delay, and duration) for each environment based on a large set of measurement data.

Purely stochastic channel models can be obtained either from geometric and numerical analysis without using measurements or they can be wholly empirical. Early cellular radio channel models, e.g., the COST 207 models, are examples of the latter. These types of models are becoming less and less common over time though, as incorporation of known physical information is shown to improve accuracy, and the greater model complexity is no longer prohibitive because of continuing advances in computer memory capacity and computational power. Geometric based channel models for AG propagation generally require three spatial dimensions to be accurate. The associated velocity vector for UAV motion in space also requires three dimensions, although 2D approximations can often be very accurate.  In order to model the scatterers around the GS, two elliptical planes intersecting a main ellipsoid were considered in~\cite{UAV_geometric1,UAV_geometric4,UAV_geometric6,UAV_geometric7}, where the MPCs are defined by the ellipsoid and the two elliptical planes. Scatterers are considered to be randomly distributed on two spheres surrounding the TX and the RX in \cite{UAV_geometric2}. In~\cite{UAV_geometric8,UAV_geometric9}, the distribution of scatterers around the GS is modeled using a three dimensional cylinder. 

\begin{table*}[!h]
	\begin{center}
      \footnotesize
		\caption{Large scale AG propagation channel fading characteristics.}\label{Table:Largescale1}
\hspace*{-.5cm}\begin{tabular}{@{}|P{2.8cm}|P{.6cm}|P{1.5cm}|P{2.6cm}|P{2.2cm}|P{2.8cm}|P{2.2cm}|@{}}
			\hline
			\textbf{Scenario}& \textbf{Ref.}& \textbf{Path (LOS/NLOS)} &\textbf{Model type} & \textbf{PLE ($\gamma$) or $(\alpha,\beta)$ parameters}& \textbf{Intercept \pmb{$PL_{\rm 0}$} (dB)} &  \pmb{$\sigma$} (dB)\\
\hline			
{Suburban, open field}&\cite{UAV_meas2}&LOS,OLOS&log-distance PL&$\gamma: 2.54-3.037$&$21.9-34.9$ &2.79-5.3  \\
\hline
{Hilly suburban}&\cite{UAV_meas6}&LOS&-&$-$&-&3.2-3.6 L-band, 1.9-3 C-band\\
\hline
{Lightly hilly rural (for $h_{\rm U}$  = $120$~m) for other values of the height, see Table~II in the paper}&\cite{Mr_Uwe1}&LOS&log-distance PL (alpha-beta model)&$\alpha = 2.0$,~$\beta = -35.3$&-&$3.4$\\
\hline
{Urban, suburban, rural}&\cite{UAV_meas7}&-&Free space PL&$-$&-&-\\
\hline
{Urban}&\cite{UAV_meas8}&-&Free space PL&$-$&-&-\\
\hline
{Urban}&\cite{UAV_meas9}&LOS&Log-distance PL&$\gamma: 1.6$ L-band, $1.9$ C-band&102.3 L-band, 113.9 C-band&-\\
\hline
{Urban, suburban}&\cite{UAV_meas12}&LOS&Log-distance PL, two ray model&$\gamma: 1.7$ L-band, $1.5-2$ C-band& $ 98.2-99.4$ L-band, $110.4-116.7$ C-band&$2.6-3.1$ L-band, $2.9-3.2$ C-band\\
\hline
{Urban, suburban}&\cite{UAV_meas14}&LOS,NLOS&Modified free space PL&-&-&-\\
\hline
{Urban, open field}&\cite{UAV_802.11_2}&LOS&Log-distance PL&$\gamma: 2.2-2.6$&-&-\\
\hline
{Urban}&\cite{UAV_meas17}&LOS,NLOS&Modified free space PL&-&-&-\\
\hline
{Urban}&\cite{UAV_meas21}&-&Modified LUI model&-&-&-\\
\hline
{Urban, rural}&\cite{AG_meas2}&LOS&Log-distance PL&$\gamma: 4.1$&-&5.24\\
\hline
{Near airports}&\cite{AG_meas4}&LOS&Log-distance PL&$\gamma: 2-2.25$&-&-\\
\hline
{Open field}&\cite{UAV_802.11_3}&-&Log-distance PL&$\gamma: 2.01$&-&-\\
\hline
{-}&\cite{UAV_802.11_4}&LOS&Log-distance PL& $\gamma: 2.32$&-&-\\
\hline
{Hilly, mountainous}&\cite{UAV_meas10}&LOS&Log-distance PL&$\gamma: 1.3-1.8$ L-band, $1-1.8$ C-band&$96.1-106.5$ L-band, $115.4-123.9$ C-band&$3.2-3.9$ L-band, $2.2-2.8$ C-band\\
\hline
{Forest/foliage}&\cite{UAV_challenges4}&-&-&-&-&-  \\
\hline
{Over sea}&\cite{UAV_meas4}&LOS&Two ray PL&-&-&-\\
\hline
{Over water, sea}&\cite{UAV_meas10}&LOS&Log-distance PL, two ray PL&$\gamma: 1.9,1.9$ over water and sea for L-band, $1.9,1.5$ over water and sea for C-band&$104.4,100.7$ over water and sea for L-band, $116.3,116.7$ over water and sea for C-band&$3.8-4.2$ over water and sea for L-band, $3.1-2.6$ for over water and sea for C-band\\
\hline
{Over sea}&\cite{AG_meas1}&LOS&Two ray PL, log distance PL, free space PL&$\gamma: .14-2.46$&$19-129$&-\\
\hline
{Ensemble of containers, see Table~II in the paper}&\cite{Mr_Uwe2}&LOS&Dual slope, &$-$&$-$&-\\
\hline

\end{tabular}
\end{center}
\end{table*}
            
The geometry-based stochastic channel models (GBSCMs) can be further classified into regular shaped GBSCMs (RS-GBSCMs) or irregular shaped GBSCMs (IS-GBSCMs). For RS-GBSCMs, the scatterers are assumed to be distributed on regular shapes e.g., ellipsoids, cylinders, or spheres. These models often result in closed form solutions, but are of course generally unrealistic. In contrast, the IS-GBSCM distributes the scatterers at random locations through some statistical distribution. The properties of the scatterers in both cases are generally defined beforehand. In some cases, authors assume a large number of scatterers a priori, and via the Central Limit Theorem, obtain a Ricean amplitude distribution to obtain estimates of the CIR based upon some geometry. Alternatively, signal interaction from randomly distributed scatterers can be estimated directly, or with the help of ray tracing software~\cite{UAV_sim2,UAV_sim4,UAV_sim16}. A non-geometric stochastic channel model (NGSCM) based on a Markov process is provided in \cite{AG_sim1}. The ground to air fading channel was described by a Markov process that switches between the Ricean and Loo models, dependent on the flight altitude.

\subsection{Path Loss and Large Scale Fading Models} \label{LSF}
As noted, in mostly-LOS AG channels, large scale fading only occurs when the LOS path between UAV and GS gets obstructed by an object that is large relative to a wavelength. Some models for this attenuation mechanism exist (e.g., terrain diffraction, tree shadowing), but not much measurement data for UAV channels obstructed by buildings has been reported. When the LOS path does not get obstructed, the only other truly large-scale effect is the two-ray variation from the earth surface MPC. There are numerous measurement campaigns in the literature for PL estimation in different environments, as summarized in Table~\ref{Table:Largescale1}. Large scale fading models in the literature cover both the PL and shadowing.

In the majority of the literature, the well-known terrestrial based log-distance PL model with free space propagation path loss reference ("close-in," CI) is used: 
\begin{align}
L_{CI}(d) = PL_{\rm 0} + 10 \gamma\log_{10}(d/d_{\rm 0}) + X_{\rm FS},\label{Eq:1}
\end{align} 
where $L_{CI}(d)$ is the model path loss as a function of distance, $PL_{\rm 0}$ is the PL at reference distance $d_0$ in free space given by $10log[\big(\frac{4\pi d_{\rm 0}}{\lambda})^2]$, $\gamma$ is the path loss exponent (PLE) obtained using minimum mean square error best fit, and $X_{\rm FS}$ is a random variable to account for shadowing, or in the case of LOS channels, the variation about the linear fit. In free space the value of PLE is 2, but as seen from Table~\ref{Table:Largescale1}, measured values of PLE  vary from approximately 1.5 to 4. One might conceptually divide the path between the UAV and the GS into two components: the free space component above the ground and the remaining \textit{terrestrial influenced} components. When the GS antenna height is well above surrounding obstacles, we expect the terrestrial components to have smaller effect and the PLE is near to that of free space. 

Another PL model used in the literature for large scale fading is floating intercept~(FI)~\cite{PL_FI}. This model is similar to (\ref{Eq:1}), but the free space PL at reference distance is eliminated and the model is dependent on two parameters represented as $\alpha$ and $\beta$~\cite{Mr_Uwe1}, where $\alpha$ is the slope and $\beta$ represents the intercept given as 
\begin{equation}
L_{\rm FI}(d) = \alpha 10\log_{10}(d) + \beta + X_{\rm FI}, \label{Eq:2}
\end{equation}
where $X_{\rm FI}$ is a random variable representing the variation of the PL.

The two PL models discussed above are based on single slope. These models hold in areas where the characteristics of the channel do not change drastically. However, in some settings with NLOS paths and complex geometries resulting in higher order reflections and diffractions, these single-slope models can have large regression errors. In such cases, a dual slope~(DS) PL model is sometimes used~\cite{PL_DS,Mr_Uwe2}. This model is similar to the FI model, but has two different slopes for different link distance ranges, and can be represented as
\begin{align}\small
&L_{\rm DS}(d) = \nonumber \\
&\begin{cases}
\alpha_{\rm d_1} 10\log_{10}(d) + \beta_{\rm d_1} + X_{\rm DS},  & d\leq d_1 \\  
\alpha_{\rm d_1} 10\log_{10}(d_1) + \beta_{\rm d_1} + \alpha_{\rm d_2} 10\log_{10}(d/d_1) + X_{\rm DS}, 
& d>d_1
\end{cases}   \label{Eq:3}
\end{align} 
where $\alpha_{d_1}$, $\alpha_{d_2}$ are the slopes of the fits for at two link distance ranges separated by threshold $d_1$, and $X_{\rm DS}$ is a random variable representing the variation in the fit.

PL estimates using log-distance models (\ref{Eq:1}) are given in~\cite{AG_meas1,AG_meas3,AG_meas4,UAV_802.11_1,UAV_802.11_2,UAV_802.11_3,UAV_802.11_4,UAV_geometric1, UAV_meas2,UAV_meas9,UAV_meas10,UAV_meas11,UAV_meas12,UAV_meas15,UAV_meas21,UAV_sim8}. There are other PL models that consider shadowing for NLOS paths, and additional losses incurred from other obstacles PL~\cite{UAV_meas14,UAV_meas8,UAV_meas17}. Due to the potential three dimensional motion of UAVs,  modified free space PL models accounting for UAV altitude can also be developed; several that are a function  of elevation angle are considered in~\cite{UAV_sim2,UAV_sim4,UAV_sim15}.

The two ray PL model described earlier in subsection~\ref{Subsection:AG_Prop_channel} is provided in~\cite{AG_meas1,Survey_Matolak,UAV_meas3,UAV_meas5,UAV_meas6,UAV_meas11,UAV_meas12,UAV_meas13,UAV_meas16}. In case of two ray PL modeling, the variation of the PL with distance has distinctive peaks due to destructive summation of the dominant and surface-reflected component. In the majority of PL models, PL variation is approximated as a log-normal random variable. This variation can be either due to shadowing from the UAV body~(see next subsection) or from MPCs attributable to terrestrial scatterers such as buildings ~\cite{AG_meas1,AG_meas2,AG_meas4,Survey_Matolak,UAV_802.11_4,UAV_geometric1,UAV_meas2,UAV_meas6,UAV_meas9,UAV_meas10,UAV_meas12,UAV_meas13,UAV_meas14}. 

In \cite{Mr_Uwe1}, log-distance FI models for the path loss exponent and shadowing for the AG radio channel between airborne UAVs and cellular networks is presented for 800 MHz and UAV heights from $1.5$~m to $120$~m above ground. In \cite{Mr_Uwe2}, the low altitude AG UAV wireless channel has been investigated for a scenario where a UAV was flying above an ensemble of containers at 5.76 GHz. Narrow- and wideband measurements have been carried out. The paper presents a modified path loss model and power delay profiles. Most interesting is that in this particular environment, delay dispersion actually increases with altitude as the UAV rises above metallic structures.

Another common model used in the literature~\cite{pathloss_new1,pathloss_new2,pathloss_new3,pathloss_new4,pathloss_new5,UAV_sim3,pathloss_new7} averages the path loss over the probabilities of LOS and NLOS path loss as follows~\cite{UAV_sim3, PL_ITU}: 
\begin{align}
{\rm PL}_{\rm avg} = P({\rm LOS})\times {\rm PL}_{\rm LOS} + \big[1-P({\rm LOS})\big]\times {\rm PL}_{\rm NLOS}~,\label{Eq:PathLoss}
\end{align}
where ${\rm PL}_{\rm LOS}$ and ${\rm PL}_{\rm NLOS}$ are the path loss in LOS and NLOS conditions, respectively,  $P({\rm LOS})$ denotes the probability of having a LOS link between the UAV and the ground node, given by~\cite{UAV_sim3,PL_ITU}:
\begin{align}
P({\rm LOS})=\prod_{n=0}^m\left[1-\exp\left(-\frac{[h_{\rm U}-\frac{(n+1/2)(h_{\rm U}-h_{\rm G})}{m+1}]^2}{2\Omega^2}\right)\right],\label{eq:P_LOS}
\end{align}
where we have $m={\rm floor}(r\sqrt{\varsigma\xi}-1)$, $r$ is the horizontal distance between the UAV and the ground node, $h_{\rm U}$ and $h_{\rm G}$ are as shown in Fig.~1 of this survey, 
$\varsigma$ is the ratio of built-up land area to the total land area, $\xi$ is the mean number of buildings per unit area (in km$^2$), and $\Omega$ characterizes the height (denoted by $H$) distribution of buildings, which is based on a Rayleigh distribution ($P(H)=(H/\Omega)^2\exp(-H/2\Omega^2)$). In~\cite{UAV_sim3}, for a specific value of $\theta$ in Fig.~2 of \cite{UAV_sim3}, a sigmoid function is also fitted to~\eqref{eq:P_LOS} for different environments (urban, suburban, dense urban, and highrise urban) to enable analytical tractability of UAV height optimization. Since~\eqref{Eq:PathLoss} averages the path loss over large number of potential LOS/NLOS link possibilities, it should be used carefully if used with system-level analysis while calculating end metrics such as throughput and outage. Similarly, path loss variability should be added to the model of~\eqref{Eq:PathLoss}.

Selection of a suitable PL model for a given AG propagation scenario is pivotal. In most of the literature, the PL model for of~(\ref{Eq:1}) is used due to its simplicity and provision of a standard platform based on reference distance free space propagation loss for comparison of measurements in different environments. A reference distance of $1$~m is often taken as a standard for short-range systems, but larger values are also used. However, in some scenarios, where the reference free space propagation loss is not available, the FI model (\ref{Eq:2}) may be used. Yet due to lack of any standard physical reference, the FI slope cannot be deemed PLE and will be dependent on the environment. Additionally, the variability of the PL is generally a zero mean Gaussian random variable that has approximately similar values for both the CI and FI model types. 

A general recommendation for selection of path loss model for a given measurement scenario from Table~\ref{Table:Largescale1} is as follows: for an open flat or hilly area with light suburban, rural or no terrain cover, and for over water, the two ray PL model or free space reference log-distance model (\ref{Eq:1}) may be preferred. For open flat or hilly environments with urban terrain cover, or for complex geometrical environments with longer NLOS paths, a dual slope PL (\ref{Eq:3}) or free space reference log-distance PL \ref{Eq:1} may be best. The FI model in (\ref{Eq:2}) may be preferred in certain specialized environments e.g., \cite{Mr_Uwe2}. In Table~\ref{Table:Largescale1}, the model types denoted log-distance refer to the general log-distance equation for path loss with different reference distances and additional parameters.

\begin{table*}[!h]
	\begin{center}
      \footnotesize
		\caption{Small scale AG propagation channel fading characteristics.}\label{Table:SmallScale1}
          \hspace*{-.5cm} \begin{tabular}{|P{.7cm}|P{2.7cm}|P{2.0cm}|P{1.2cm}|P{1.9cm}|P{1.0cm}|P{2.2cm}|P{3.0cm}|}
\hline
\textbf{Ref.}&\textbf{Scenario} & \textbf{Time-variant/Time-invariant} & \textbf{Modeling type}& \textbf{Frequency spectrum}& \textbf{DS (Hz)}&  \textbf{Fading distribution}& \textbf{\textit{K}-factor (dB)}\\
\hline
\cite{UAV_meas1}&{Urban/Suburban}&Time-invariant&Statistical&Narrow-band&-&Ricean&- \\
\hline
\cite{UAV_meas2}&{Suburban/Open field}&Time-invariant&Statistical&Ultra-wideband&-&Nakagami&- \\
\hline
\cite{UAV_meas7}&{Suburban/Open field}&Time-variant&-&-&833&-&- \\
\hline
\cite{UAV_meas8}&{Urban/Suburban}&-&-&Narrow-band&&-&- \\
\hline
\cite{AG_meas2}&{Urban/suburban}&Time-invariant&Statistical&Wide-band&-&Rayleigh, Ricean&-\\
\hline
\cite{AG_meas4}&{Urban/suburban}&-&Statistical&Wide-band&$1400$&Ricean&(-$5$)-$10$\\
\hline
\cite{UAV_meas12}&{Urban/Suburban}&Time-variant&Statistical&Wide-band&-&Ricean&$12$-$27.4$ in L and C band \\
\hline
\cite{AG_meas5}&{Hilly}&Time-variant&-&Wide-band&$10000$&-&-\\
\hline
\cite{UAV_challenges4}&{Forest/foliage}&-&Statistical&Ultra-wideband&-&Ricean, Nakagami&$2$-$5$\\
\hline
\cite{UAV_meas11}&{Sea/fresh water}&Time-variant&Statistical&Wide-band&-&Ricean&$12$, $28$ for L and C band\\
\hline
\cite{AG_meas7}&{-}&Time-variant&Statistical&Wide-band&$5820$&-&-\\
\hline
		\end{tabular}
		\end{center}
			\end{table*}

\subsection{Airframe Shadowing}
Airframe shadowing occurs when the body of the aircraft itself obstructs the LOS to the GS. This impairment is somewhat unique to AG communications, and not much exists in the literature on this effect. One reason for this is that such shadowing can be largely (but not always completely) alleviated by using multiple spatially separated antennas: airframe shadowing on one antenna can be made unlikely to occur at the same time as shadowing on the other(s). In addition to frequency and antenna placement, shadowing results also depend on the exact shape, size, and material of the aircraft. For small rotorcraft, depending on frequency and antenna placement, airframe shadowing could be minimal. Example measurement results, as well as models for airframe shadowing, for a fixed wing medium sized aircraft, were provided in \cite{AF_shadowing_Matolak}.

For these results, at frequencies of 970 and 5060 MHz, wing shadowing attenuations were generally proportional to aircraft roll angle, with maximum shadowing depths exceeding 35 dB at both frequencies. Shadowing durations depend upon flight maneuvers, but for long, slow banking turns, can exceed tens of seconds.

\begin{figure}[!t]
	\centering
	\includegraphics[width=\columnwidth]{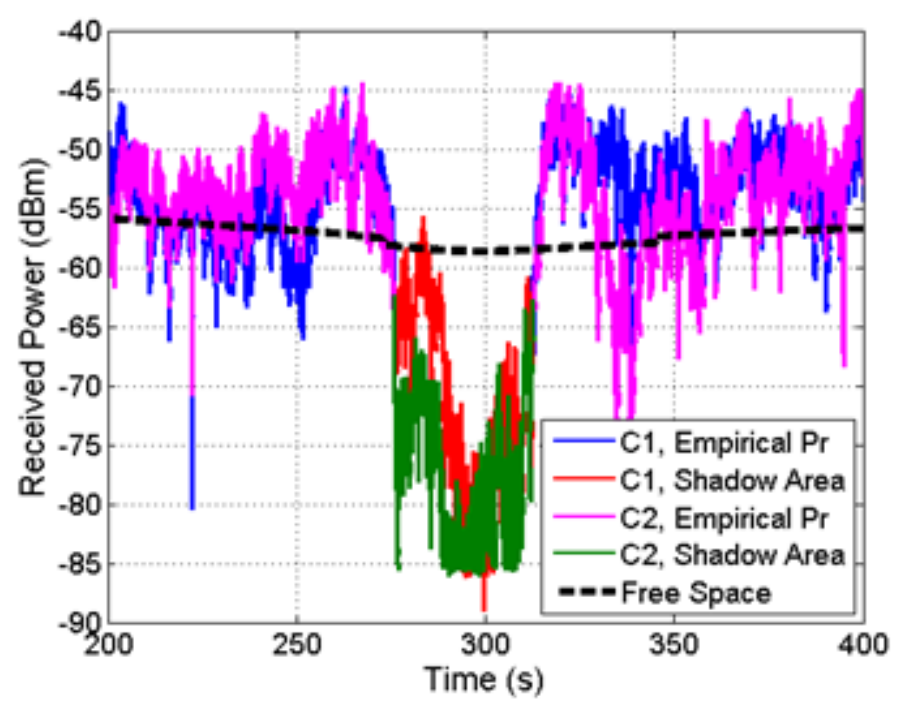}
	\caption{Received power vs. time for illustration of shadowing before, during and after the banking turn of medium sized aircraft at C-band.}\label{Fig:Air_frame_shadowing}
\end{figure}

An illustration of airframe shadowing is shown in Fig.~\ref{Fig:Air_frame_shadowing} where received power is plotted against time for a wideband ($50$~MHz) signal in C-band before, during, and after the medium-sized aircraft made a banking turn. The received power on two aircraft antennas (denoted C1, C2), bottom mounted and separated by approximately $1.2$~m, is shown. Attenuations due to airframe shadowing, along with the polarization mismatch that occurs during the aircraft maneuver, exceed approximately $30$~dB in this case.  

\subsection{Small Scale Fading Models}
Small-scale fading models apply to narrow-band channels or to individual MPCs, or \textit{taps} in tapped delay line wide-band models, with bandwidth up to some maximum value (i.e., small scale fading may not pertain to MPCs in a UWB channel). The depth of small scale amplitude fades on a given signal also generally varies inversely with signal bandwidth ~\cite{BW_SSfading}. Stochastic fading models are obtained through analysis, empirical data, or through geometric analysis and simulations~\cite{UAV_geometric1,UAV_geometric2,UAV_geometric4,UAV_geometric6,UAV_geometric8,UAV_geometric9}. As noted in subsection~\ref{Subsection:Channel_models}, the GBSCMs can be subdivided into RS-GBSCM and IS-GBSCM. In~\cite{UAV_geometric6}, a time-variant IS-GBSCM was provided with a Ricean distribution for small scale fading. Time-variant RS-GBSCM were provided in ~\cite{UAV_geometric2,UAV_geometric8}, and these also illustrated Ricean small scale fading. 

A NGSCM was provided in \cite{AG_sim1}, where GA fading was described using Ricean and Loo models. The Loo model was derived based on the assumption that the amplitude attenuation of the LOS component due to foliage in a land mobile satellite link follows a log-normal distribution, and that the fading due to MPCs follows a Rayleigh distribution. The switching between Ricean and Loo models was controlled by a Markov process dependent on flight height. In~\cite{UAV_geometric4}, a GBSCM for MPCs was provided in the form of shape factors describing angular spread, angular compression, and direction of maximum fading using the probability density function (PDF) of angle of arrival.

Table~\ref{Table:SmallScale1} provides measured small scale AG fading characteristics reported in the literature for various environments. As previously noted, the most common small scale fading distribution for AG propagation is the Ricean. As in terrestrial channels, for the NLOS case, the Rayleigh fading distribution typically provides a better fit~\cite{AG_geometric2,AG_meas2,AG_meas4,UAV_geometric5,UAV_meas1,UAV_sim1,UAV_sim9}, and of course, other distributions such as the Nakagami-m and Weibull distributions might also be employed. Small scale fading rates depend upon velocity, and these rates are proportional to the Doppler spreads of the MPCs ~\cite{UAV_meas7,AG_meas4,UAV_meas5,UAV_meas7,UAV_sim14,AG_sim1}. 



            
\subsection{Intermittent MPCs}
Another AG characteristic that may be of interest in high-fidelity and long-term channel models is the intermittent nature of MPCs. From geometry, it is easy to deduce that for a given vehicle trajectory in some environment, individual MPCs will persist only for some finite span of time ~\cite{UAV_meas13}. This has been noted in V2V channels as well, but with UAVs and their potentially larger velocities, the intermittent MPC (IMPC) dynamics can be greater. These IMPCs arise (are "born") and disappear ("die") naturally in GBSCMs. They may also be modeled using discrete time Markov chains. The IMPCs can significantly change the CIR for some short time span, hence yielding wide variation in RMS-DS. (Another manifestation of so-called "non-stationarity.") Example models for the IMPCs--their probability of occurrence, duration, delay, and amplitude--appear in ~\cite{UAV_meas4,UAV_meas13,UAV_meas16}.

\begin{figure}[!t]
	\centering
	\includegraphics[width=\columnwidth]{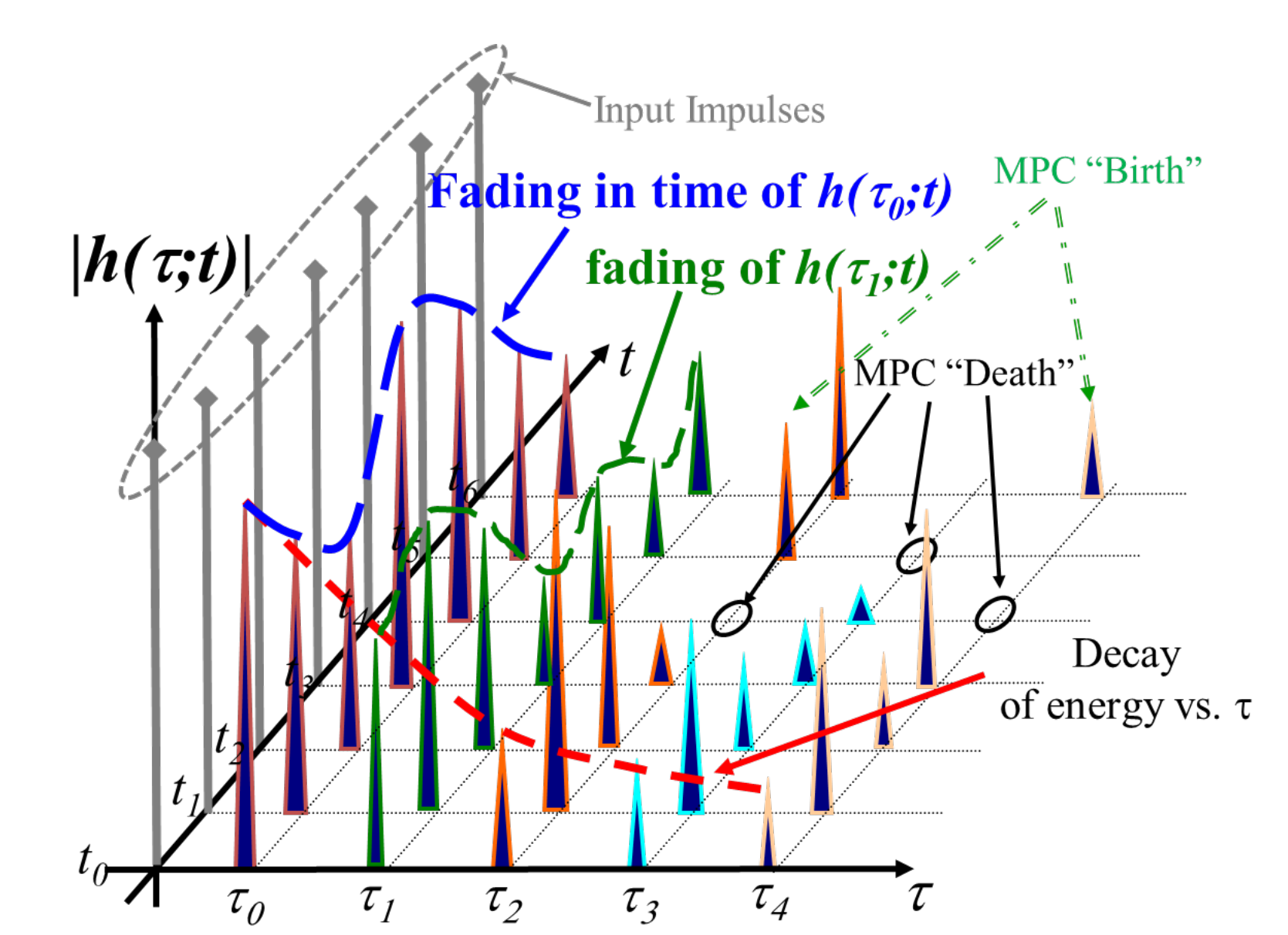}
	\caption{Fading and birth and death process of intermittent MPCs from \cite{Survey_Matolak}.}\label{Fig:MPCs_birth_death}
\end{figure}

In Fig.~\ref{Fig:MPCs_birth_death}, from \cite{Survey_Matolak} the fading of MPCs as a function of time and delay are shown. The amplitude of MPCs generally decay with excess delay at a given time instant. Additionally, there is a continuous birth and death process of MPCs at different instants of time. This can be represented using CIR as~\cite{Survey_Matolak}:
\begin{equation}
h(t,\tau) = \sum_{i=0}^{M(t)-1}p_{i}(t)a_{i}(t)\exp(j\phi_{i}(t))\delta(\tau - \tau_{i}(t)) ,\label{Eq:CIR_Matolak}
\end{equation}
where $h(t,\tau)$ is the time variant channel impulse response, $M(t)$ is the total number of MPCs at time instant $t$, $p_{i}(t)$ represents the multipath persistence process coefficient and can take binary values $[0,1]$. The amplitude, phase and delay of $i^{\rm th}$ MPC at time instant $t$ are represented as $a_{i}(t)$, $\phi_{i}(t)$ and $\tau_{i}(t)$ respectively. The phase term is given as $\phi_{i}(t) = 2\pi{f_{d}^{i}(t)(t-\tau_{i}(t)) - f_{\rm c}(t)\tau_{i}(t)}$, where  $f_{d}^{i}(t) = v(t)f_{\rm c}(t)\cos(\Theta_{i}(t)/c$ is the Doppler frequency of the $i^{\rm th}$ MPC, $\Theta_{i}(t)$ is the aggregate phase angle in the $i^{\rm th}$ delay bin, $c$ is the speed of light and $f_{\rm c}$ represents the carrier frequency.
The channel transfer function $H(f,t)$ from (\ref{Eq:CIR_Matolak}) is then given as follows: 
\begin{equation}
\begin{split}
H(f,t) &= \sum_{i=0}^{M(t)-1}p_{i}(t)a_{i}(t)\exp\Big(j2\pi f_{d}^{i}(t)\big(t-\tau_{i}(t)\big)\Big)\\
&\times\exp\big(-j2\pi f_{\rm c}\tau_{i}(t)\big)\exp\big(-j2\pi f\tau_{i}(t)\big), \label{Eq:CTF_Matolak}
\end{split}
\end{equation}

The effect of the Doppler spread is typically negligible compared to carrier frequency at lower velocities. Therefore the carrier frequency term will dominate the variation of the transfer function.

\begin{figure}[!t]
	\centering
	\includegraphics[width=\columnwidth]{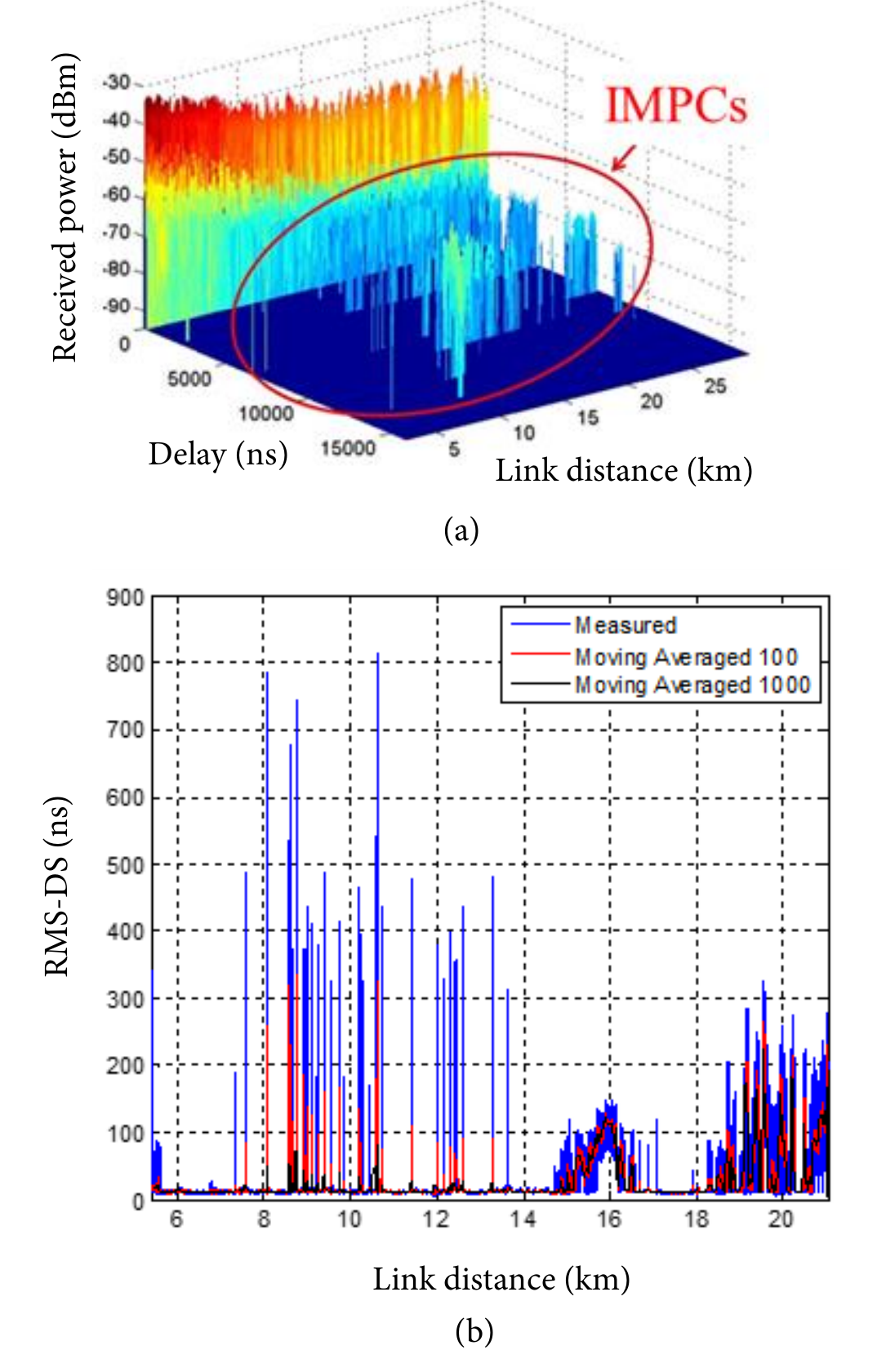}
	\caption{(a) Sequence of PDPs versus link distance for a near-urban AG link near Cleveland, OH, (b) RMS-DS vs. link distance for a hilly environment in Palmdale, California. }\label{Fig:Intermittent_MPCs}
\end{figure}

Fig.~\ref{Fig:Intermittent_MPCs}(a) shows a sequence of PDPs versus link distance for a near-urban AG link near Cleveland, OH. Flight parameters can be found in \cite{UAV_meas12}. In this figure, the IMPCs are clearly visible, here caused by reflections from obstacles near the Lake Erie shoreline. In Fig.~\ref{Fig:Intermittent_MPCs}(b) RMS-DS vs. link distance for a hilly environment in Palmdale, California is shown. The intermittent nature of the MPCs produces “spikes” and “bumps” in the RMS-DS values, illustrating the potential rapid time variation of AG channels.
 
\subsection{MIMO AG Propagation Channel Models}
The use of MIMO systems for AG UAV communications has been gaining popularity. The rationale, increased throughput and reliability, is the same one driving mmWave and future 5G research. In~\cite{MIMO_antenna}, it was shown that it is possible to attain higher spatial multiplexing gains in LOS channels by properly selecting the antenna separation and orientation as a function of carrier wavelength and link distance. This careful alignment is not always practical or possible with UAVs, especially when mobile.

The advantages of spatial diversity and multiplexing gains in MIMO are often only moderate due to limited scattering available near UAVs or GSs. In~\cite{UAV_sim_mimo1}, it was demonstrated that due to limited spatial diversity in the AG channel, only moderate capacity gains are possible. In order to obtain better spatial multiplexing gains, larger antenna separations are required, and this requires large antenna arrays that are not feasible on-board small UAVs. Use of higher carrier frequencies makes it possible to use electrically-large antenna arrays, but higher frequencies yield higher PL~(this can be mitigated somewhat by beamforming, at the expense of the complexity required for beam steering). Moreover, accurate channel state information~(CSI) is important for MIMO systems for higher performance, but in a rapidly varying AG propagation channel, it can be difficult to provide accurate CSI and hence MIMO gains can be limited. The use of MIMO on airborne platforms also incurs additional cost, computational complexity, and power consumption.

There is a limited number of studies available in the literature for MIMO AG propagation channel measurements. In~\cite{UAV_meas_mimo2}, a detailed measurement analysis of the AG MIMO propagation channel was provided. It was observed that a considerable spatial de-correlation of the received signal at the GS is achieved due to the interaction of non-planar wavefronts. These wavefronts are generated due to near field effects from the measurement vehicle, on which the GS antennas were mounted. Spatial diversity from antennas located on the UAV was also observed, interestingly at \textit{higher} elevation angles. The authors suggest that having scatterers near the GS can yield larger spatial diversity. The received signal in~\cite{UAV_meas18} was analyzed for multiple-input-single-output~(MISO) and MIMO systems, and it was observed that the use of MIMO systems enables a more robust channel for changes in antenna orientations arising from UAV maneuvering. In~\cite{UAV_meas20}, MIMO system performance was tested in different scenarios of the outdoor environment, including urban, rural, open field, and forest. The effect of terrain cover on the received power was analyzed for these different scenarios with the result that the propagation channel in the open field is mostly influenced by the ground reflections, whereas in case of forests, the reflection and shadowing from the trees is a major contributor to the propagation channel characteristics. In rural and urban cases, the reflections from the walls and surfaces of building structures play an important role.

Time-variant GBSCMs for MIMO systems provided in \cite{UAV_geometric6,UAV_geometric3,UAV_geometric8,UAV_geometric2} were explored through simulations. A simulation based AG MIMO channel propagation model was provided in~\cite{UAV_geometric1} for a hilly area. The results indicate increased throughput from spatial multiplexing and higher SNR from the MIMO system in comparison to SISO, as expected. A stochastic model for a mobile to mobile AG MIMO propagation channel was presented in~\cite{UAV_geometric2}.  These results show that there was considerable capacity increase and reduction in outage probabilities using MIMO systems if perfect instantaneous CSI is available. In~\cite{UAV_geometric3}, geometry-based simulations were conducted for a massive MIMO implementation for a UAV AG propagation channel. The simulation results illustrate the expected result of a significant capacity increase when a large number of antennas is used at the GS.

\subsection{Ray Tracing Simulations} 
\begin{figure}[!t]
	\centering
	\includegraphics[width=\columnwidth]{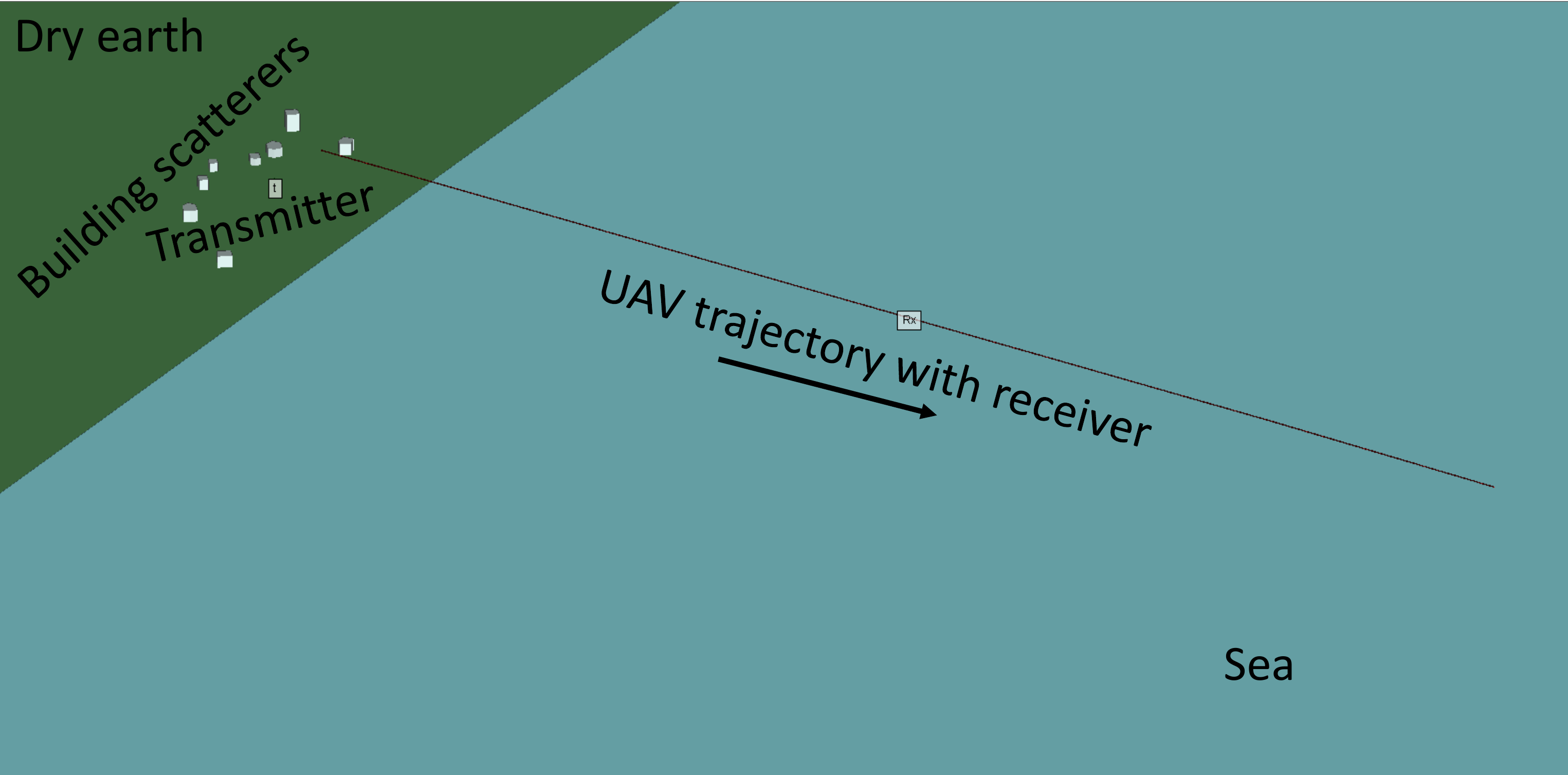}
	\caption{Ray tracing simulation scenario for over sea scenario, where the UAV flies over a straight line.}\label{Fig:Ray_tracing}
\end{figure}

\begin{figure}[!t]
	\centering
	\includegraphics[width=\columnwidth]{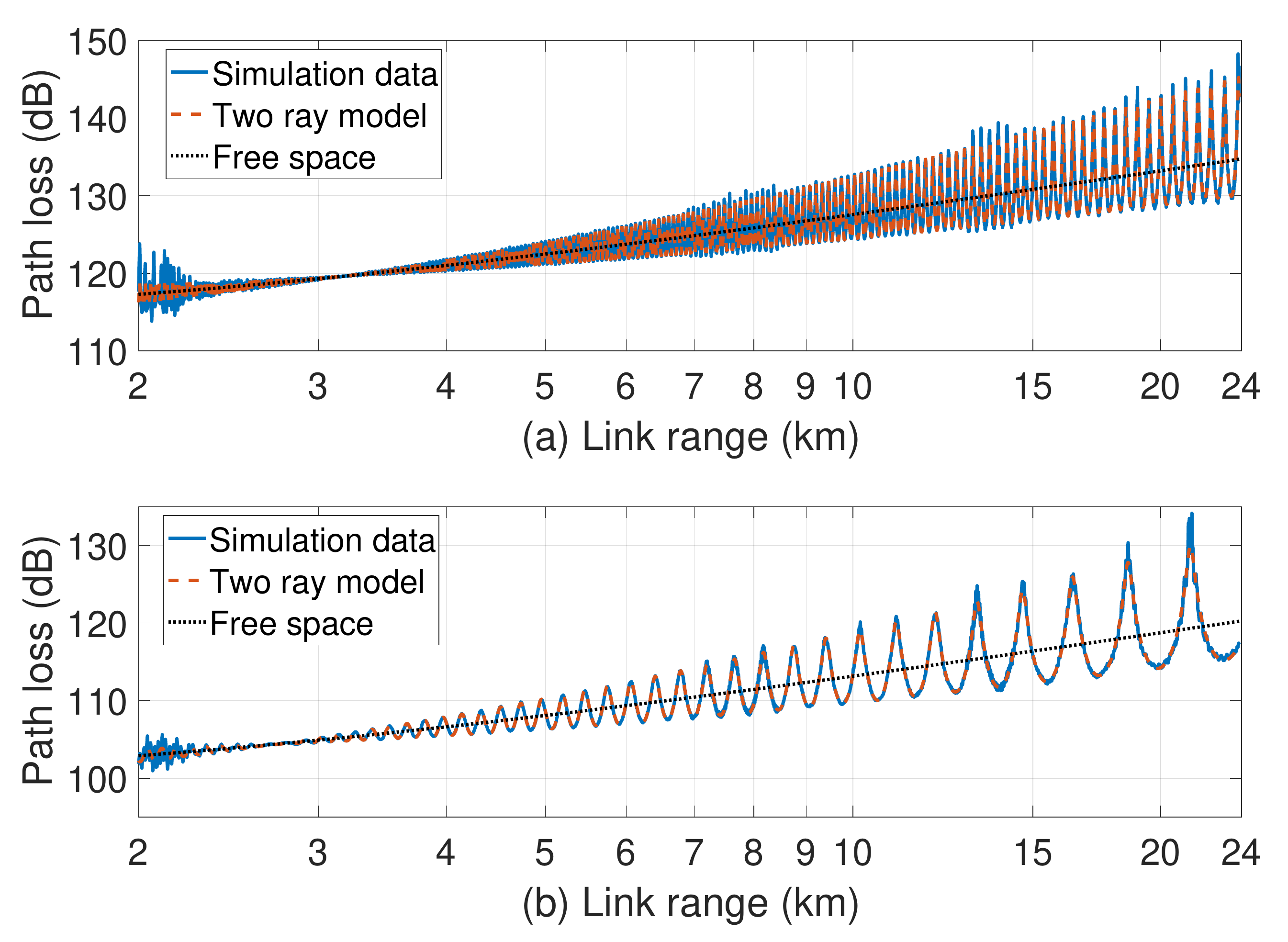}
	\caption{Ray tracing PL results for over sea water settings, (a) C-band~(5.03~GHz - 5.091~GHz) , (b) L-band~(0.9~GHz - 1.2~GHz).}\label{Fig:Sea_PL}
\end{figure}

\begin{figure}[!t]
	\centering
	\includegraphics[width=\columnwidth]{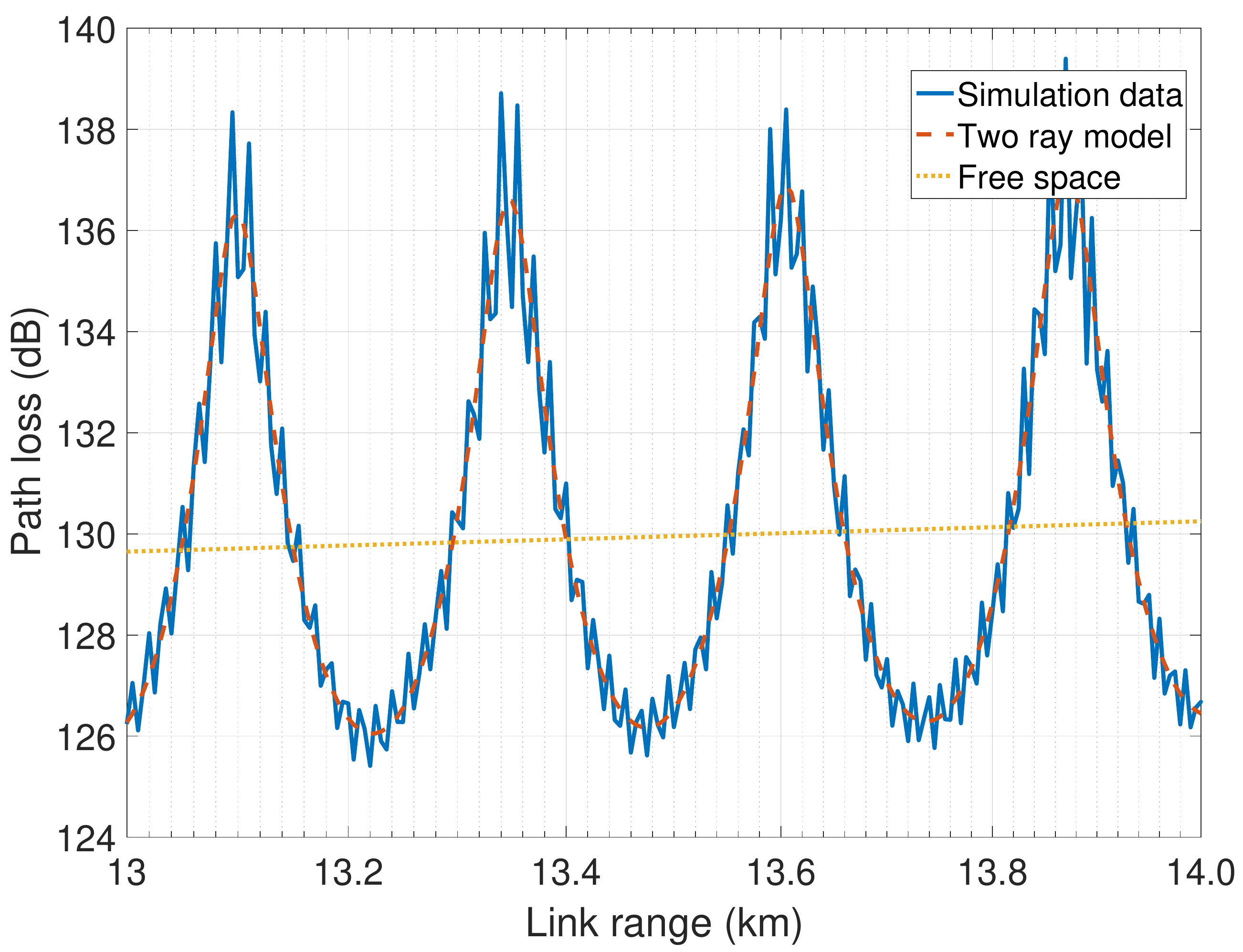}
	\caption{Zoomed in results of PL for over sea water simulations of Fig.~\ref{Fig:Sea_PL} for C-band at link distances of 
    $13$~km - $14$~km.}\label{Fig:Sea_PL_zoom}
\end{figure}

\begin{figure}[t!]
    \centering
    \subfigure[]{
        \includegraphics[width=\columnwidth]{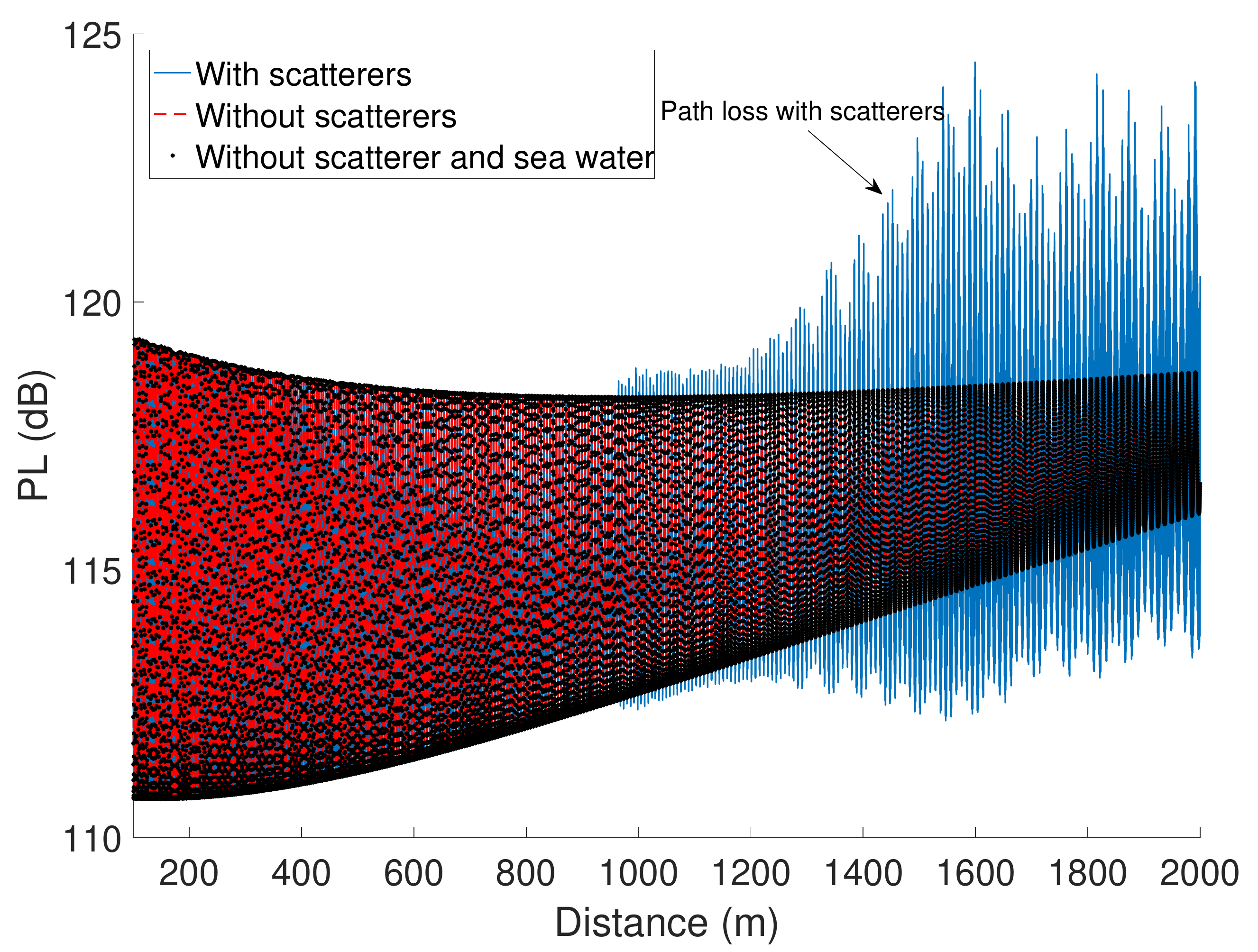}
    }
    ~ 
    \subfigure[]{
        \includegraphics[width=\columnwidth]{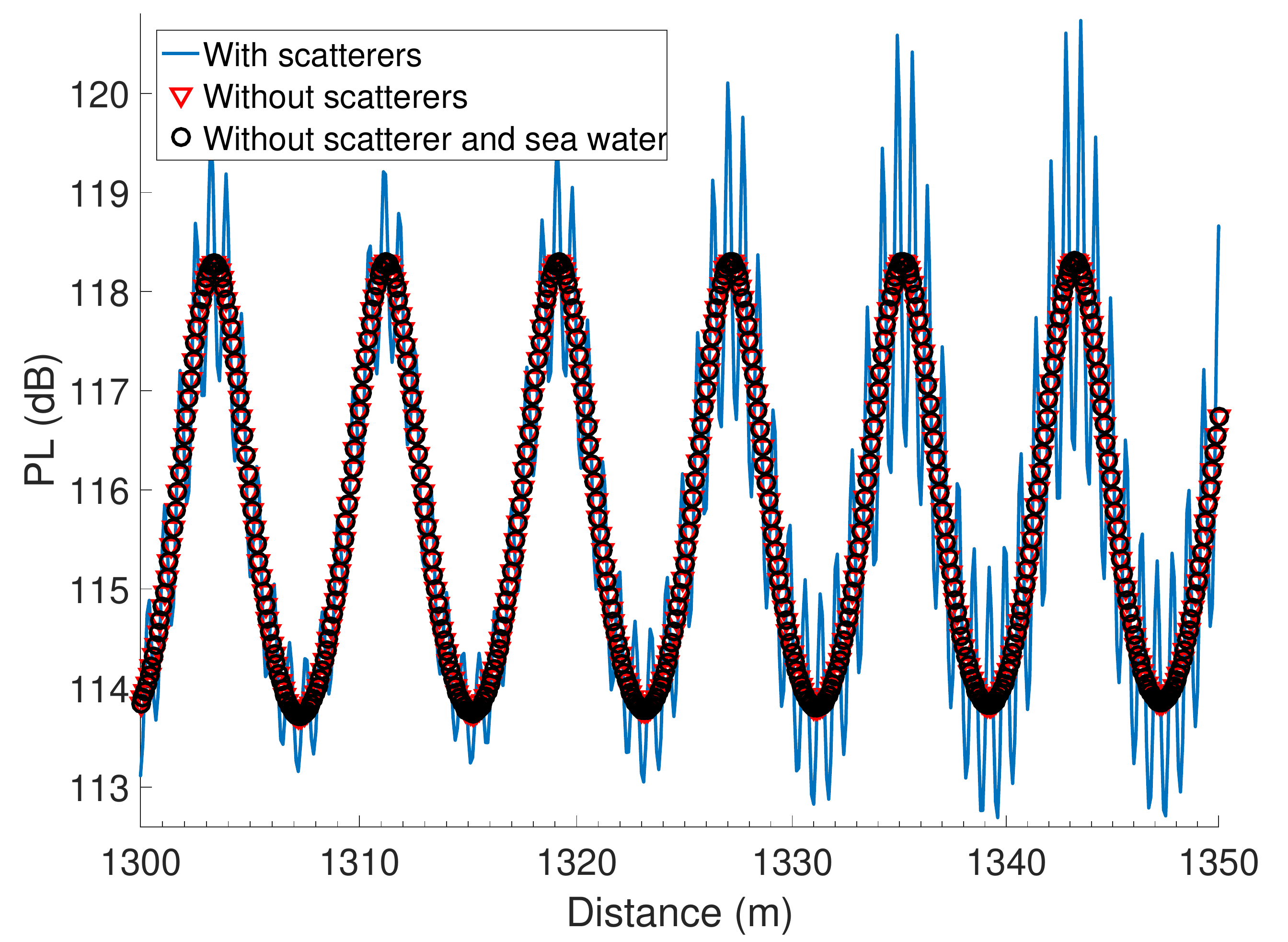}
	}
\caption{Path loss versus distance with/without scatterers, and without the sea surface: (a) $100$~m to $2$~km range, and (b) $1300$~m to $1350$~m range.}\label{Fig:LongDistancePL}
\end{figure}

\begin{figure}[!t]
	\centering
	\includegraphics[width=\columnwidth]{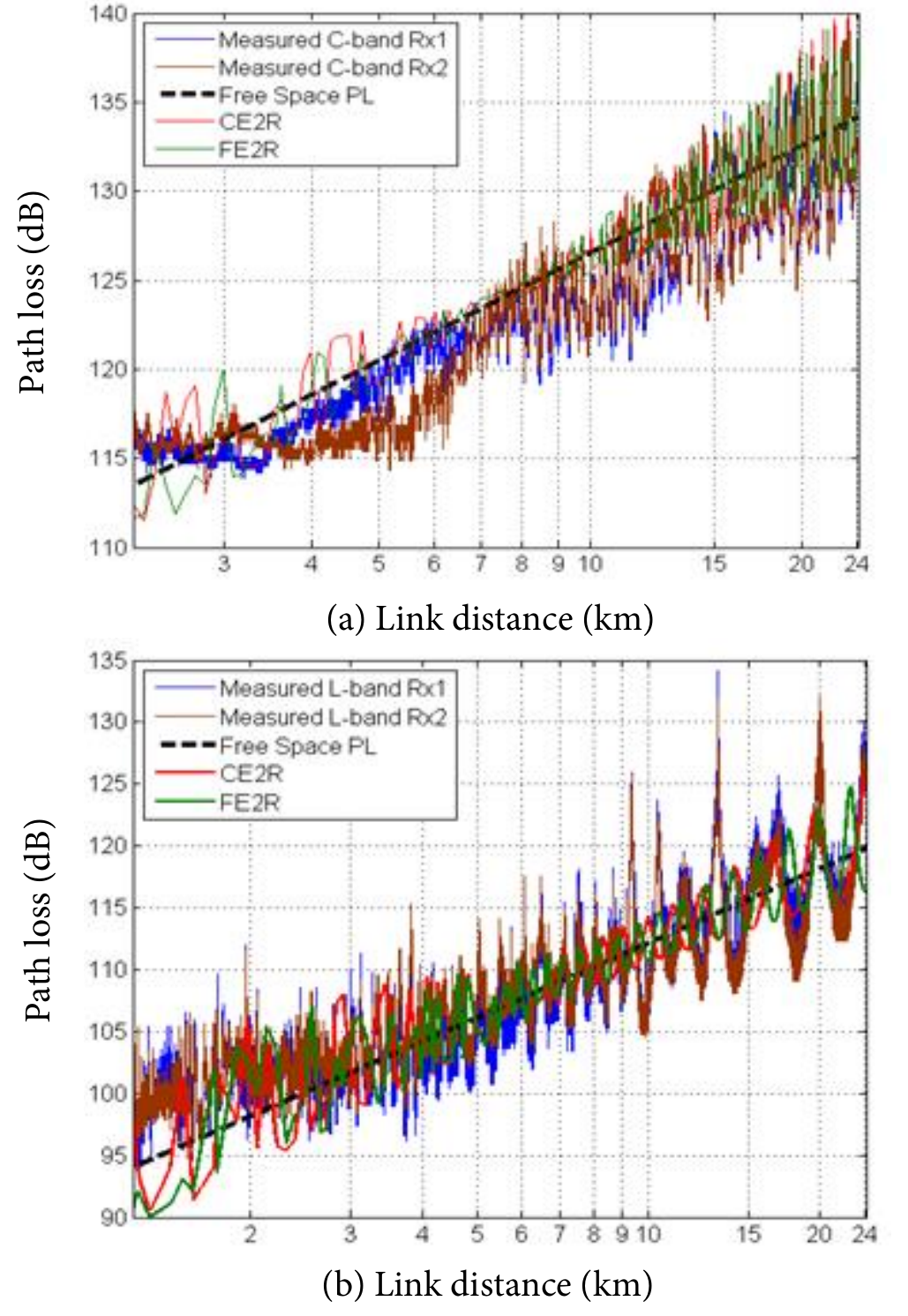}
	\caption{Measurement results for PL over sea scenario from \cite{UAV_meas13}: (a) C-band~(5.03~GHz - 5.091~GHz), (b) L-band~(0.9~GHz - 1.2~GHz).}\label{Fig:Oversea_meas_PL}
\end{figure}
In the literature, in addition to measurements, channel characterization for AG propagation is also carried out using simulations. These simulators are either based on customized channel environments on a given software platform or can be realized using ray tracing simulations. There are PL models available for these simulated environments~\cite{UAV_sim1,UAV_sim2,UAV_sim3,UAV_sim4,UAV_sim8,UAV_sim15}. Urban environmental scenarios for LOS and NLOS paths were considered in~\cite{UAV_sim4,UAV_sim2,UAV_sim3} where log-distance and modified free space PL models were suggested. In~\cite{UAV_sim8} a log-distance path model was provided for LOS and NLOS paths for over sea settings in a simulated environment. However, to the best knowledge of the authors, there are no specific experimental studies available in the literature that experimentally validate the channel models proposed using geometrical analysis and simulations in ~\cite{UAV_sim1,UAV_sim2,UAV_sim3,UAV_sim4,UAV_sim8,UAV_sim15}.

Ray tracing was used for mmWave channel characterization for 28~GHz and 60~GHz frequency bands for UAV AG propagation in \cite{wahab_mmWave}. Different environments were realized, namely urban, suburban, rural and over sea. It was observed that the RSS follows that of the two ray model and is further affected by the presence of scatterers in the surroundings. The RMS-DS was also affected by the presence of scatterers in the surrounding environment and the UAV height in the given environment. If the height of the scatterers is large with dimensions large relative to a wavelength, we observe higher RMS-DS for higher UAV altitudes due to multiple reflections from the densely distributed scatterers. In contrast, if the height of the scatterers is small, we have smaller RMS-DS at higher UAV heights due to fewer significant MPCs reaching the UAV. This phenomena is verified at $28$~GHz and $60$~GHz, where at $60$~GHz, we have smaller RMS-DS than at $28$~GHz due to higher attenuation of MPCs.

Ray tracing simulations using Wireless InSite software were carried out to estimate PL for an over-sea scenario as shown in Fig.~\ref{Fig:Ray_tracing}. The channel measurement parameters were set according to \cite{UAV_meas13}, and the simulated PL results were compared with the measured values. Fig.~\ref{Fig:Sea_PL} shows the simulated PL results. In this simulated environment, we have buildings as scatterers near the transmitter. 
Due to reflections and diffractions from these scatterers we observe additional fluctuations on top of the two ray propagation model. 
The deviations 
are due to MPCs reflected and diffracted from the different-shaped scatterers at different angles. These weak MPCs reach the UAV receiver at different link distances resulting in  variations from the two ray model as shown in Fig.~\ref{Fig:Sea_PL_zoom} at a link distance between $13$~km-$14$~km.

Similarly in Fig.~\ref{Fig:LongDistancePL}(a), the effect of MPCs from scatterers around the TX for link distances 100~m-2~km are shown. It can be observed that without the scatterers and seawater~(with ground only), we have a perfect two ray PL model. Yet in the presence of the scatterers around the TX, superimposed upon this effect is variation from additional MPCs from the scatterers; this yields what can be modeled as a random path loss component on top of the two ray model, or in effect a small scale fading. This effect is of course dependent on the geometry of the scenario and will cause the path loss to vary  along the trajectory of the UAV. A similar effect at the larger link distance range of $13$~km-$13.5$~km  in Fig.~\ref{Fig:LongDistancePL}(a) can be observed in Fig.~\ref{Fig:LongDistancePL}(b).


Fig.~\ref{Fig:Oversea_meas_PL} shows measured and model PL results from \cite{UAV_meas13} for over-sea settings, where CE2R and FE2R stands for curved earth two ray and flat earth two ray model, respectively. There is a good match between the ray tracing simulation results and analytical results for this over sea scenario in Fig.~\ref{Fig:Sea_PL}, but when comparing measurement data with simulation data, we observe more fluctuations in measurements due to several factors: measurement equipment variation, ambient noise, and in particular scattering from the rough sea surface, which is not as easily modeled with the basic ray tracing. Also plotted along with the measurement data in Fig.~\ref{Fig:Oversea_meas_PL} are analytical results for free space and curved- and flat-earth two ray models. The curved- and flat-earth two-ray models are obtained using the specific geometry and conditions of the measurement environment.   

\section{Future Research Areas for AG UAV Channel Measurements and Models}
 In this section we discuss limitations of currently available AG channel measurements and models and their possible enhancements. We also identify some representative considerations for future research. Our aim is to incite development of more comprehensive, realistic, and accurate propagation channel models for future UAV communication applications.

\subsection{Future small UAV scenarios}

In future scenarios small UAVs will fly in cities, across suburban areas, and over rural terrain. There are two conceptually very different communication approaches for small UAVs: the first approach is based on centralized communications, i.e., UAVs communicate with base stations similar to the concept of 3G and 4G cellular mobile radio. These base stations would preferably be located on elevated positions such as towers or roof tops and have antennas whose radiation patterns are optimized for serving these UAVs. The second approach foresees direct communications among all UAVs, similar to vehicular communications such as ITS-G5 (intelligent transportation systems communications standard at 5.9 GHz). Both approaches have their pros and cons in terms of robustness, latency, and capacity; as implied, no decision has been made so far on which approach to use and only a few channel measurements have been carried out so far for both approaches. 

The scenarios that have to be considered for future propagation measurements should encompass urban, suburban, industrial, rural, and even indoor or "quasi-confined" areas such as large arenas or stadiums. Attention should be directed not only to en-route situations; even though these might be less demanding in terms of propagation conditions, strong multipath components are likely to occur due to reflections from smooth wet ground or bodies of water, and from large buildings with metallized window fronts. In addition we also recommend investigating the channel for take-off and landing scenarios, be it on roof tops, in gardens, or in other specifically assigned areas. In these take-off and landing conditions, propagation may be unfavorable due to shadowing, strong diffraction, and rich multipath, and it is in these cases where communication must work very reliably. Moreover, we think that the propagation conditions for flights that bring UAVs intentionally close to building facades, power lines, containers, and other objects (e.g., for inspection) should also be investigated as such propagation may exhibit special or atypical features. 

\subsection{UAV AG Propagation Measurements}
Existing AG propagation channel measurements and models mostly apply to aeronautical communications at higher flight altitudes than envisioned for small UAVs. These smaller structures have limited on-board computation capabilities, strict power limitations, and can only fly at much lower altitudes, and at present, only for short durations. There is a growing demand for higher data rates, low latency, and high reliability for future communications, and this will be challenging for current civilian UAV architectures. 

Additionally, as noted in Section~\ref{Section:Introduction}, there are usually two types of communications maintained simultaneously for UAVs: payload and CNPC. However, currently there are no standards adopted worldwide for these two types of communications for UAVs. Both can have their own operating bands that may or may not overlap. The CNPC communication links are pivotal for maintaining safety of flight and any interference can be catastrophic. Standards organizations are thus working on robust \textit{loss of link} procedures. Moreover, the CNPC needs to be secure and resistant to jamming and hacking attacks. The USA has developed a standard, primarily for medium and large aircraft~\cite{CNPC_standard}, with standards envisioned for smaller UAVs in the future. 

Future measurement campaigns should take into account not only a great variety of buildings - small and large ones, rectangular and irregularly shaped ones, industrial facilities, halls, and towers - but also reflecting areas like bodies of water, streets, and squares, and demanding situations when a UAV lands on a terrace or the like. Especially for modeling the UAV-to-UAV channel, different velocities and flight situations should be investigated, e.g., two UAVs flying toward each other, with one UAV near ground and the other up in the air, and swarms of UAVs flying with the same velocity. For cellular-like deployments, interference is likely to be a significant issue that influences network planning. Thus, it would be useful to have measurements up to far distancess (and over different terrain). We envisage that the UAV-to-UAV channel for small UAVs in urban areas is as diverse as the car-to-car channel, the latter being modeled as a 2.5 dimensional channel whereas the UAV-to-UAV channel will often need to be modeled as a 3 dimensional channel.

In addition to the UAV settings, there are several other factors that need to be taken into account for comprehensive AG propagation measurements using UAVs. One of these is the placement and orientation of antennas. The placement of antennas should be such that there is minimum shadowing and noise effect from the air-frame and motors while flying. Achieving this is not always easy, and will usually be UAV-specific. The antenna orientation has been shown to result in different throughputs and RSS values~\cite{UAV_802.11_1,UAV_802.11_2,UAV_802.11_3,UAV_802.11_4} for different flight maneuvers. In order to provide better coverage during flight, omni-directional antennas on both TX and RX are commonly used, especially for CNPC communications. The use of directional antennas is dependent on the specific application and coverage. When selecting UAV antenna options, the mechanical viability for a given UAV type should also be taken into account e.g., a long helical antenna or yagi uda structure may be difficult to mount on a fixed wing aircraft compared to a horn or patch antenna.

There is no fixed number of antennas recommended for optimum performance, and the number of antennas will depend on the operating frequency, UAV size, and operational environments. In many experiments multiple antennas are used on UAVs, and these may be helpful for improved coverage and diversity gains, but at the expense of increased computation, space, and power requirements. 

The ambient conditions on-board UAVs must also be taken into account for precise measurement of any communication link characteristics~(for CNPC or otherwise). These ambient conditions include noise from the motors, noise from aircraft electronics, air friction while moving, sudden air gusts, temperature variations, and outside-system interference. The latter may be particularly severe for unlicensed bands. Another consideration with the use of unlicensed bands and commodity radios is that the adaptive modulation and coding algorithms employed for terrestrial networks (which often assume quasi-stationary conditions) may not work so well when directly applied to highly dynamic UAV AG propagation channels. 

Nearly all current day channel measurements take advantage of positioning information, typically from global navigation satellite systems, with GPS being the most widely used. In addition to position information, GPS signals also provide an accurate time reference. Depending on measurement requirements and the envisioned application, the accuracy of GPS may or may not be sufficient, and this should be considered before beginning measurement campaigns.

When using UAVs in swarms, the location and mobility aware routing methods that are used for terrestrial networks may need to be adapted to account for the three dimensional movement of UAVs. Similarly, route selection algorithms for mobility aware networks will need to consider the fast varying channel conditions during UAV flight.  

\subsection{UAV AG Propagation Channel Models}
The UAV AG propagation literature mostly covers the modeling of PL, as described in Section~\ref{LSF}. As noted, and as is common for terrestrial channels, the PL models are typically provided as a function of link distance. For UAVs there might be other models appropriate for certain cases, for example a PL model as a function of UAV altitude in a given setting, or even indoor UAV PL models for certain settings (e.g., large arenas).

The most accurate UAV AG propagation channel models are of course time varying, but in some cases these can be specialized to time-invariant approximations, e.g., when a UAV is hovering above an area of static objects. In~\cite{UAV_meas4,UAV_meas10,UAV_meas13,UAV_meas16}, the channel is considered to be quasi stationary only for short distances, and small scale fading parameters are evaluated over that stationarity interval. Additional studies of the stationarity distance should be conducted for other UAV propagation scenarios, using multiple metrics: the PDP correlation coefficient, correlation matrix collinearity, spectral divergence, and evolutionary spectrum have all been used, but each metric has its own advantages and disadvantages. Depending on environments, additional UAV measurement campaigns will likely result in more elaborate UAV AG propagation channel models, that may make use of MPC clusters, spatial (angular) information, and correlations among model parameters. Ultimately, deterministic and hybrid channel models using GBSCM principles will likely evolve to be the most widely used to characterize UAV AG propagation.

\section{Concluding Remarks}
In this paper, we have provided a comprehensive survey for AG propagation channels for UAVs. The measurement campaigns in the literature for AG propagation were summarized, with information provided on the type of channel sounding signal, its center frequency, bandwidth, transmit power, UAV speed, height of UAV and GS, link distance, elevation angle, and local GS environment characteristics. Air-ground channel statistics from the literature were also provided. Various UAV propagation scenarios and important implementation factors for these measurements were also discussed. Large scale fading, small scale fading, MIMO channel characteristics and models, and channel simulations were all described. Finally, future research directions and challenges were discussed. We expect that more elaborate and accurate AG propagation measurement campaigns and channel models will be developed in the future, and we hope this study will be of use in that regard.
 
\bibliographystyle{IEEEtran}

\begin{thebibliography}{100}
\providecommand{\url}[1]{#1}
\csname url@samestyle\endcsname
\providecommand{\newblock}{\relax}
\providecommand{\bibinfo}[2]{#2}
\providecommand{\BIBentrySTDinterwordspacing}{\spaceskip=0pt\relax}
\providecommand{\BIBentryALTinterwordstretchfactor}{4}
\providecommand{\BIBentryALTinterwordspacing}{\spaceskip=\fontdimen2\font plus
\BIBentryALTinterwordstretchfactor\fontdimen3\font minus
  \fontdimen4\font\relax}
\providecommand{\BIBforeignlanguage}[2]{{%
\expandafter\ifx\csname l@#1\endcsname\relax
\typeout{** WARNING: IEEEtran.bst: No hyphenation pattern has been}%
\typeout{** loaded for the language `#1'. Using the pattern for}%
\typeout{** the default language instead.}%
\else
\language=\csname l@#1\endcsname
\fi
#2}}
\providecommand{\BIBdecl}{\relax}
\BIBdecl

\bibitem{Tractica}
{Tractica}, ``Commercial drone shipments to surpass 2.6 million units annually
  by 2025,'' accessed: 2017-11-28.

\bibitem{Drones}
\BIBentryALTinterwordspacing
{Wikipedia}, ``General atomics {MQ}-9 {R}eaper,'' accessed: 2017-07-03.
  [Online]. Available:
  \url{https://en.wikipedia.org/wiki/General_Atomics_MQ-9_Reaper}
\BIBentrySTDinterwordspacing

\bibitem{FAA_new}
\BIBentryALTinterwordspacing
{Federal Aviation Administration}, ``{FAA} small unmanned aircraft
  regulations,'' accessed: 2017-07-03. [Online]. Available:
  \url{https://www.faa.gov/news/fact_sheets/news_story.cfm?newsId=20516}
\BIBentrySTDinterwordspacing

\bibitem{Qual}
\BIBentryALTinterwordspacing
{Qualcomm}, ``Leading the world to 5{G}: {E}volving cellular technologies for
  safer drone operation,'' accessed: 2017-05-17. [Online]. Available:
  \url{https://www.qualcomm.com/invention/technologies/lte/advanced-pro/cellular-drone-communication}
\BIBentrySTDinterwordspacing

\bibitem{CNN}
\BIBentryALTinterwordspacing
T.~Patterson, ``{Google}, {F}acebook, {SpaceX}, {OneWeb} plan to beam internet
  everywhere,'' accessed: 2017-05-17. [Online]. Available:
  \url{http://www.cnn.com/2015/10/30/tech/pioneers-google-facebook-spacex-oneweb-satellite-drone-balloon-internet/}
\BIBentrySTDinterwordspacing

\bibitem{history1}
F.~White, ``Air-ground communications: history and expectations,'' \emph{{IEEE
  Trans. Commun.}}, vol.~21, no.~5, pp. 398--407, May 1973.

\bibitem{history3}
M.~S.~B. Mahmoud, C.~Guerber, A.~Pirovano, N.~Larrieu \emph{et~al.},
  \emph{Aeronautical Air-Ground Data Link Communications}.\hskip 1em plus 0.5em
  minus 0.4em\relax John Wiley \& Sons, 2014.

\bibitem{manned_2}
M.~Schnell, U.~Epple, D.~Shutin, and N.~Schneckenburger, ``{LDACS}: future
  aeronautical communications for air-traffic management,'' \emph{{IEEE Commun.
  Mag.}}, vol.~52, no.~5, pp. 104--110, May 2014.

\bibitem{militaryAG}
\BIBentryALTinterwordspacing
{RadioReference.com}, ``{VHF}/{UHF} military monitoring,'' accessed:
  2017-05-31. [Online]. Available:
  \url{http://wiki.radioreference.com/index.php/VHF/UHF_Military_Monitoring}
\BIBentrySTDinterwordspacing

\bibitem{manned_1}
E.~Haas, ``Aeronautical channel modeling,'' \emph{{IEEE Trans. Vehic.
  Technol.}}, vol.~51, no.~2, pp. 254--264, Mar. 2002.

\bibitem{aeronautical1}
B.~G. Gates, ``Aeronautical communications,'' \emph{Electrical Engineers - Part
  {IIIA}: Radiocommunication, Journal}, vol.~94, no.~11, pp. 74--81, Mar. 1947.

\bibitem{aeronautical2}
D.~F. Lamiano, K.~H. Leung, L.~C. Monticone, W.~J. Wilson, and B.~Phillips,
  ``Digital broadband {VHF} aeronautical communications for air traffic
  control,'' in \emph{Proc. Integrated Communs., Navigation and Surveillance
  Conf.}, May 2009, pp. 1--12.

\bibitem{Survey_Matolak}
D.~W. Matolak, ``Air-ground channels models: Comprehensive review and
  considerations for unmanned aircraft systems,'' in \emph{{Proc. IEEE
  Aerospace Conf.}}, Mar. 2012, pp. 1--17.

\bibitem{HAPs_satellites}
C.~Levis, J.~T. Johnson, and F.~L. Teixeira, \emph{Radiowave propagation:
  physics and applications}.\hskip 1em plus 0.5em minus 0.4em\relax John Wiley
  \& Sons, 2010.

\bibitem{HAP1}
G.~M. Djuknic, J.~Freidenfelds, and Y.~Okunev, ``Establishing wireless
  communications services via high-altitude aeronautical platforms: a concept
  whose time has come?'' \emph{{IEEE Commun. Mag.}}, vol.~35, no.~9, pp.
  128--135, Sep. 1997.

\bibitem{CNPC1}
B.~Kerczewski, ``Spectrum for {UAS} control and non-payload communications,''
  in \emph{{Proc. IEEE Integrated Communs., Navigation and Surveillance Conf.
  (ICNS)}}, 2013, pp. 1--21.

\bibitem{CNPC2}
B.~R. Jackson, ``Telemetry, command and control of {UAS} in the {N}ational
  {A}irspace,'' in \emph{{Proc. Int. Telemetering Conf.}}\hskip 1em plus 0.5em
  minus 0.4em\relax Int. Foundation for Telemetering, 2015.

\bibitem{UAV_challenges1}
M.~Asadpour, B.~V. den Bergh, D.~Giustiniano, K.~A. Hummel, S.~Pollin, and
  B.~Plattner, ``Micro aerial vehicle networks: an experimental analysis of
  challenges and opportunities,'' \emph{{IEEE Commun. Mag.}}, vol.~52, no.~7,
  pp. 141--149, Jul. 2014.

\bibitem{UAV_challenges2}
Y.~Zeng, R.~Zhang, and T.~J. Lim, ``Wireless communications with unmanned
  aerial vehicles: opportunities and challenges,'' \emph{{IEEE Commun. Mag.}},
  vol.~54, no.~5, pp. 36--42, May 2016.

\bibitem{UAV_challenges3}
L.~Afonso, N.~Souto, P.~Sebastiao, M.~Ribeiro, T.~Tavares, and R.~Marinheiro,
  ``Cellular for the skies: Exploiting mobile network infrastructure for low
  altitude air-to-ground communications,'' \emph{{IEEE Aerospace and Electronic
  Systems Mag.}}, vol.~31, no.~8, pp. 4--11, Aug. 2016.

\bibitem{UAV_Specs3}
Z.~Xiao, P.~Xia, and X.~g.~Xia, ``Enabling {UAV} cellular with millimeter-wave
  communication: potentials and approaches,'' \emph{{IEEE Commun. Mag.}},
  vol.~54, no.~5, pp. 66--73, May 2016.

\bibitem{UAV_doppler1}
M.~Ibrahim and H.~Arslan, ``Air-ground doppler-delay spread spectrum for dense
  scattering environments,'' in \emph{{ Proc. IEEE Military Commun. Conf.
  MILCOM}}, Oct. 2015, pp. 1661--1666.

\bibitem{UAV_doppler2}
R.~Essaadali and A.~Kouki, ``A new simple unmanned aerial vehicle doppler
  effect {RF} reducing technique,'' in \emph{{Proc. IEEE Military Commun.
  Conf.}}, Nov. 2016, pp. 1179--1183.

\bibitem{UAV_sim3}
A.~Al-Hourani, S.~Kandeepan, and S.~Lardner, ``Optimal {LAP} altitude for
  maximum coverage,'' \emph{{IEEE Wireless Commun. Letters}}, vol.~3, no.~6,
  pp. 569--572, 2014.

\bibitem{UAV_geometric9}
L.~Zeng, X.~Cheng, C.~X. Wang, and X.~Yin, ``Second order statistics of
  non-isotropic {UAV} rician fading channel,'' in \emph{{Proc. IEEE Vehic
  Technol. Conf. (VTC)}}, Sep., 2017.

\bibitem{UAV_geometric2}
A.~Ksendzov, ``A geometrical {3D} multi-cluster mobile-to-mobile {MIMO} channel
  model with {R}ician correlated fading,'' in \emph{{Proc. IEEE Int. Congress
  Ultra Modern Telecommuns. (ICUMT) Conf.}}, 2016, pp. 191--195.

\bibitem{UAV_geometric4}
S.~M. Gulfam, S.~J. Nawaz, A.~Ahmed, and M.~N. Patwary, ``Analysis on multipath
  shape factors of air-to-ground radio communication channels,'' in
  \emph{{Proc. IEEE Wireless Telecomm. Symposium (WTS)}}, 2016, pp. 1--5.

\bibitem{UAV_geometric7}
S.~M. Gulfam, J.~Syed, M.~N. Patwary, and M.~Abdel-Maguid, ``On the spatial
  characterization of 3-{D} air-to-ground radio communication channels,'' in
  \emph{{Proc. IEEE Int. Conf. Commun. (ICC)}}, 2015, pp. 2924--2930.

\bibitem{UAV_meas9}
D.~W. Matolak and R.~Sun, ``Air-ground channel characterization for unmanned
  aircraft systems: The near-urban environment,'' in \emph{{Proc. IEEE Military
  Commun. Conf., MILCOM}}, 2015, pp. 1656--1660.

\bibitem{UAV_meas1}
M.~Simunek, F.~P. Fontán, and P.~Pechac, ``The {UAV} low elevation propagation
  channel in urban areas: Statistical analysis and time-series generator,''
  \emph{{IEEE Trans. Ant. Propag.}}, vol.~61, no.~7, pp. 3850--3858, Jul. 2013.

\bibitem{UAV_meas13}
{D. W. Matolak and R. Sun}, ``Air ground channel characterization for unmanned
  aircraft systems part {I}: Methods, measurements, and models for over-water
  settings,'' \emph{{IEEE Trans. Vehic. Technol.}}, vol.~66, no.~1, pp. 26--44,
  Jan. 2017.

\bibitem{Antenna_stat1}
S.~Kaul, K.~Ramachandran, P.~Shankar, S.~Oh, M.~Gruteser, I.~Seskar, and
  T.~Nadeem, ``Effect of antenna placement and diversity on vehic. network
  communications,'' in \emph{{Proc. IEEE Sensor, Mesh and Ad Hoc Communs. and
  Networks}}, Jun. 2007, pp. 112--121.

\bibitem{Antenna_stat2}
A.~R. Ruddle, ``Simulation of far-field characteristics and measurement
  techniques for vehicle-mounted antennas,'' in \emph{IEE Colloquium on
  Antennas for Automotives (Ref. No. 2000/002)}, 2000, pp. 7/1--7/8.

\bibitem{AG_meas2}
W.~G. Newhall, R.~Mostafa, C.~Dietrich, C.~R. Anderson, K.~Dietze, G.~Joshi,
  and J.~H. Reed, ``Wideband air-to-ground radio channel measurements using an
  antenna array at 2 {GH}z for low-altitude operations,'' in \emph{{Proc. IEEE
  Military Commun. Conf. (MILCOM)}}, vol.~2, 2003, pp. 1422--1427.

\bibitem{UAV_meas4}
{D. W. Matolak and R. Sun}, ``Antenna and frequency diversity in the unmanned
  aircraft systems bands for the over-sea setting,'' in \emph{Proc. IEEE
  Digital Avionics Sys. Conf. (DASC)}, Oct. 2014, pp. 6A4--1--6A4--10.

\bibitem{UAV_meas_mimo1}
J.~Chen, B.~Daneshrad, and W.~Zhu, ``{MIMO} performance evaluation for airborne
  wireless communication systems,'' in \emph{{Proc. Military Commun. Conf.
  (MILCOM)}}, Nov. 2011, pp. 1827--1832.

\bibitem{UAV_802.11_1}
C.-M. Cheng, P.~H. Hsiao, H.~Kung, and D.~Vlah, ``Performance measurement of
  802.11 a wireless links from {UAV} to ground nodes with various antenna
  orientations,'' in \emph{{Proc. Int. Conf. Computer Communs. and Networks,
  (ICCCN)}}, 2006, pp. 303--308.

\bibitem{UAV_802.11_2}
E.~Yanmaz, R.~Kuschnig, and C.~Bettstetter, ``Channel measurements over
  802.11a-based {UAV}-to-ground links,'' in \emph{{Proc. IEEE GLOBECOM
  Workshops (GC Wkshps)}}, 2011, pp. 1280--1284.

\bibitem{UAV_meas_mimo2}
T.~J. Willink, C.~C. Squires, G.~W.~K. Colman, and M.~T. Muccio, ``Measurement
  and characterization of low-altitude air-to-ground {MIMO} channels,''
  \emph{{IEEE Trans. Vehic. Technol.}}, vol.~65, no.~4, pp. 2637--2648, Apr.
  2016.

\bibitem{SIMO}
{D. W. Matolak, H. Jamal and R. Sun}, ``Spatial and frequency correlations in
  two-ray {SIMO} channels,'' in \emph{{Proc. IEEE Int. Conf. on Communs.
  (ICC)}}, {May, 2017}.

\bibitem{UAV_geometric8}
X.~Gao, Z.~Chen, and Y.~Hu, ``Analysis of unmanned aerial vehicle {MIMO}
  channel capacity based on aircraft attitude,'' \emph{{WSEAS Trans. Inform.
  Sci. Appl}}, vol.~10, pp. 58--67, 2013.

\bibitem{AG_geometric2}
C.~Zhang and Y.~Hui, ``Broadband air-to-ground communications with adaptive
  {MIMO} datalinks,'' in \emph{{Proc. IEEE Digital Avionics Sys. Conf.
  (DASC)}}, 2011, pp. 4D4--1.

\bibitem{AG_sim1}
J.~Yang, P.~Liu, and H.~Mao, ``Model and simulation of narrowband ground-to-air
  fading channel based on {M}arkov process,'' in \emph{{Proc. Network Computing
  and Information Security Conf. (NCIS)}}, vol.~1, 2011, pp. 142--146.

\bibitem{AG_meas7}
C.~Bluemm, C.~Heller, B.~Fourestie, and R.~Weigel, ``Air-to-ground channel
  characterization for {OFDM} communication in {C}-band,'' in \emph{{Proc. Int.
  Conf. Signal Processing Commun. Sys. (ICSPCS)}}, 2013, pp. 1--8.

\bibitem{UAV_sim9}
V.~Vahidi and E.~Saberinia, ``Orthogonal frequency division multiplexing and
  channel models for payload communications of unmanned aerial systems,'' in
  \emph{{Proc. Int. Conf. Unmanned Aircraft Systems (ICUAS)}}, 2016, pp.
  1156--1161.

\bibitem{UAV_sim14}
Z.~Wu, H.~Kumar, and A.~Davari, ``Performance evaluation of {OFDM} transmission
  in {UAV} wireless communication,'' in \emph{{Proc. Southeastern Symposium
  Sys. Theory, (SSST)}}, 2005, pp. 6--10.

\bibitem{AG_meas4}
H.~D. Tu and S.~Shimamoto, ``A proposal of wide-band air-to-ground
  communication at airports employing 5-{GH}z band,'' in \emph{{Proc. IEEE
  Wireless Commun. Networking Conf. (WCNC)}}, 2009, pp. 1--6.

\bibitem{UAV_meas2}
W.~Khawaja, I.~Guvenc, and D.~W. Matolak, ``{UWB} channel sounding and modeling
  for {UAV} air-to-ground propagation channels,'' in \emph{{Proc. IEEE Global
  Commun. Conf. (GLOBECOM)}}, Dec. 2016, pp. 1--7.

\bibitem{UAV_meas3}
D.~W. Matolak and R.~Sun, ``Air-ground channel measurements \& modeling for
  {UAS},'' in \emph{{Proc. Integrated Commun., Navigation and Surveillance
  Conf. (ICNS)}}, 2013, pp. 1--9.

\bibitem{UAV_meas5}
R.~Sun and D.~W. Matolak, ``Over-harbor channel modeling with directional
  ground station antennas for the air-ground channel,'' in \emph{{Proc. IEEE
  Military Commun. Conf. (MILCOM)}}, 2014, pp. 382--387.

\bibitem{UAV_meas6}
D.~W. Matolak and R.~Sun, ``Air-ground channel characterization for unmanned
  aircraft systems: The hilly suburban environment,'' in \emph{{Proc. Vehic.
  Technol. Conf. (VTC)}}, 2014, pp. 1--5.

\bibitem{UAV_meas10}
R.~Sun and D.~W. Matolak, ``Air-ground channel characterization for unmanned
  aircraft systems—part {II}: Hilly \& mountainous settings,'' \emph{{IEEE
  Trans. Vehic. Technol.}}, 2016.

\bibitem{UAV_meas11}
D.~W. Matolak and R.~Sun, ``Air-ground channels for {UAS}: Summary of
  measurements and models for {L}-and {C}-bands,'' in \emph{{Proc. Integrated
  Commun. Navigation and Surveillance (ICNS)}}, 2016, pp. 8B2--1.

\bibitem{UAV_meas12}
{D. W. Matolak and R. Sun}, ``Air-ground channel characterization for unmanned
  aircraft systems—part {III}: The suburban and near-urban environments,''
  \emph{{IEEE Trans. Vehic. Technol.}}, 2017.

\bibitem{UAV_meas16}
D.~W. Matolak, ``Channel characterization for unmanned aircraft systems,'' in
  \emph{{Proc. European Conf. Ant. Propag. (EuCAP)}}, 2015, pp. 1--5.

\bibitem{Schneckenburger_TAES2015}
N.~Schneckenburger, T.~Jost, D.~Shutin, M.~Walter, T.~Thiasiriphet, M.~Schnell,
  and U.~Fiebig, ``Measurement of the {L}-band air-to-ground channel for
  positioning applications,'' \emph{IEEE Trans. Aerosp. Electron. Syst.},
  vol.~52, no.~5, pp. 2281--2297, Oct. 2016.

\bibitem{UAV_meas7}
K.~Takizawa, T.~Kagawa, S.~Lin, F.~Ono, H.~Tsuji, and R.~Miura, ``C-band
  aircraft-to-ground {(A2G)} radio channel measurement for unmanned aircraft
  systems,'' in \emph{{Proc. Wireless Personal Multimedia Commun. (WPMC)
  Conf.}}, 2014, pp. 754--758.

\bibitem{UAV_meas8}
F.~Ono, K.~Takizawa, H.~Tsuji, and R.~Miura, ``S-band radio propagation
  characteristics in urban environment for unmanned aircraft systems,'' in
  \emph{{Proc. Ant. Propag. (ISAP) Conf.}}, 2015, pp. 1--4.

\bibitem{UAV_meas15}
N.~Goddemeier, K.~Daniel, and C.~Wietfeld, ``Coverage evaluation of wireless
  networks for unmanned aerial systems,'' in \emph{{Proc. IEEE GLOBECOM
  Workshops (GC Wkshps)}}, 2010, pp. 1760--1765.

\bibitem{UAV_802.11_3}
E.~Yanmaz, R.~Kuschnig, and C.~Bettstetter, ``Achieving air-ground
  communications in 802.11 networks with three-dimensional aerial mobility,''
  in \emph{{Proc. IEEE INFOCOM}}, 2013, pp. 120--124.

\bibitem{UAV_802.11_4}
N.~Ahmed, S.~S. Kanhere, and S.~Jha, ``On the importance of link
  characterization for aerial wireless sensor networks,'' \emph{{IEEE Commun.
  Mag.}}, vol.~54, no.~5, pp. 52--57, 2016.

\bibitem{UAV_challenges4}
E.~L. Cid, A.~V. Alejos, and M.~G. Sanchez, ``Signaling through scattered
  vegetation: Empirical loss modeling for low elevation angle satellite paths
  obstructed by isolated thin trees,'' \emph{IEEE Vehic. Technol. Mag.},
  vol.~11, no.~3, pp. 22--28, Sep. 2016.

\bibitem{UAV_meas18}
J.~Romeu, A.~Aguasca, J.~Alonso, S.~Blanch, and R.~R. Martins, ``Small {UAV}
  radiocommunication channel characterization,'' in \emph{{Proc. European Conf.
  Ant. Propag. (EuCAP)}}, 2010, pp. 1--5.

\bibitem{UAV_meas19}
H.~Kung, C.-K. Lin, T.-H. Lin, S.~J. Tarsa, and D.~Vlah, ``Measuring diversity
  on a low-altitude {UAV} in a ground-to-air wireless 802.11 mesh network,'' in
  \emph{{Proc. IEEE GLOBECOM Workshops (GC Wkshps)}}, 2010, pp. 1799--1804.

\bibitem{UAV_meas20}
J.~Zelen{\`y}, F.~P{\'e}rez-Font{\'a}n, and P.~Pecha{\v{c}}, ``Initial results
  from a measurement campaign for low elevation angle links in different
  environments,'' in \emph{{Proc. European Conf. Ant. Propag. (EuCAP)}}, 2015,
  pp. 1--4.

\bibitem{UAV_meas22}
E.~Teng, J.~Falcao, C.~Dominguez, F.~Mokaya, P.~Zhang, and B.~Iannucci,
  ``Aerial sensing and characterization of three-dimensional {RF} fields,''
  \emph{{Univ. at Buffalo, cse. buffalo. edu, accessed: Sep}}, 2016.

\bibitem{AG_meas1}
Y.~S. Meng and Y.~H. Lee, ``Measurements and characterizations of air-to-ground
  channel over sea surface at {C}-band with low airborne altitudes,''
  \emph{{IEEE Trans. Vehic. Technol.}}, vol.~60, no.~4, pp. 1943--1948, 2011.

\bibitem{AG_meas5}
J.~Kunisch, I.~De~La~Torre, A.~Winkelmann, M.~Eube, and T.~Fuss, ``Wideband
  time-variant air-to-ground radio channel measurements at 5 {GH}z,'' in
  \emph{{Proc. European Conf. Ant. Propag. (EUCAP)}}, 2011, pp. 1386--1390.

\bibitem{wahab_mmWave}
W.~Khawaja, O.~Ozdemir, and I.~Guvenc, ``{UAV} air-to-ground channel
  characterization for mmwave systems,'' in \emph{{Proc. IEEE Vehic. Technol.
  Conf. (VTC) Sep. 2017}}, 2017.

\bibitem{air_air}
M.~Walter, S.~Gligorević, T.~Detert, and M.~Schnell, ``{UHF}/{VHF} air-to-air
  propagation measurements,'' in \emph{Proceedings of the Fourth European
  Conference on Antennas and Propagation}, Apr. 2010, pp. 1--5.

\bibitem{UAV_air_to_air}
N.~Goddemeier and C.~Wietfeld, ``Investigation of air-to-air channel
  characteristics and a {UAV} specific extension to the {R}ice model,'' in
  \emph{{Proc. IEEE Globecom Workshops (GC Wkshps)}}, Dec. 2015, pp. 1--5.

\bibitem{FAA}
\BIBentryALTinterwordspacing
{Federal Aviation Administration}, ``{FAA} rules for {UAVs},'' accessed:
  2017-02-25. [Online]. Available:
  \url{https://www.faa.gov/uas/beyond_the_basics/}
\BIBentrySTDinterwordspacing

\bibitem{VNA}
\BIBentryALTinterwordspacing
{Universitat Politècnica de Catalunya}, ``Vector network analyzer
  specifications,'' accessed: 2017-05-18. [Online]. Available:
  \url{http://www.upc.edu/sct/en/documents_equipament/d_160_id-655-2.pdf}
\BIBentrySTDinterwordspacing

\bibitem{Cover_terrain}
\BIBentryALTinterwordspacing
{International Telecommunication Union}, ``Terrain cover types,'' accessed:
  2017-07-05. [Online]. Available: \url{https://www.itu.int/oth/R0A04000031/en}
\BIBentrySTDinterwordspacing

\bibitem{ITU_mountain}
\BIBentryALTinterwordspacing
{International Telecommunication Union }, ``Propagation by diffraction,''
  accessed: 2017-07-05. [Online]. Available:
  \url{http://www.itu.int/dms_pubrec/itu-r/rec/p/R-REC-P.526-13-201311-I!!PDF-E.pdf}
\BIBentrySTDinterwordspacing

\bibitem{forest_satellite1}
M.~Kvicera, F.~P. Font{\'a}n, J.~Israel, and P.~Pechac, ``A new model for
  scattering from tree canopies based on physical optics and multiple
  scattering theory,'' \emph{{IEEE Trans. Ant. and Propag.}}, vol.~65, no.~4,
  pp. 1925--1933, 2017.

\bibitem{forest_satellite2}
F.~Kawamata, ``Optimum frame size for land mobile satellite communication
  channels,'' in \emph{Proc. IEEE Global Telecommun. Conf.(GLOBECOM)}, Nov.
  1993, pp. 583--587 vol.1.

\bibitem{forest_satellite3}
A.~A. Aboudebra, K.~Tanaka, T.~Wakabayashi, S.~Yamamoto, and H.~Wakana,
  ``Signal fading in land-mobile satellite communication systems: statistical
  characteristics of data measured in japan using {ETS-VI},'' \emph{Proc. IEEE
  Microwaves, Ant. and Propag.}, vol. 146, no.~5, pp. 349--354, Oct. 1999.

\bibitem{Duct_ITU}
\BIBentryALTinterwordspacing
{International Telecommunication Union}, ``Ducting over sea calculation,''
  accessed: 2017-07-05. [Online]. Available:
  \url{http://www.itu.int/md/dologin_md.asp?id=R03-WRC03-C-0025!A27-L188!MSW-E}
\BIBentrySTDinterwordspacing

\bibitem{ofdm_sounding}
Q.~Chen, ``Wideband channel sounding techniques for dynamic spectrum access
  networks,'' Ph.D. dissertation, University of Kansas, 2009.

\bibitem{parsons}
J.~Parsons, ``The mobile radio propagation channel vol. 2nd west sussex,''
  \emph{UK: John Wiley \& Sons, Ltd}, 2000.

\bibitem{WSS_dist1}
A.~Paier, T.~Zemen, L.~Bernado, G.~Matz, J.~Karedal, N.~Czink, C.~Dumard,
  F.~Tufvesson, A.~F. Molisch, and C.~F. Mecklenbrauker, ``Non-{WSSUS}
  vehicular channel characterization in highway and urban scenarios at 5.2
  {GH}z using the local scattering function,'' in \emph{Proc. Int. Workshop on
  Smart Antennas}, Feb. 2008, pp. 9--15.

\bibitem{WSS_dist2}
O.~Renaudin, V.~M. Kolmonen, P.~Vainikainen, and C.~Oestges, ``Non-stationary
  narrowband {MIMO} inter-vehicle channel characterization in the 5-{GH}z
  band,'' \emph{IEEE Trans. on Vehic. Technol.}, vol.~59, no.~4, pp.
  2007--2015, May 2010.

\bibitem{UAV_geometric1}
M.~Wentz and M.~Stojanovic, ``A {MIMO} radio channel model for low-altitude
  air-to-ground communication systems,'' in \emph{{Proc. IEEE Vehic. Technol.
  Conf. (VTC)}}, 2015, pp. 1--6.

\bibitem{UAV_sim2}
A.~Al-Hourani, S.~Kandeepan, and A.~Jamalipour, ``Modeling air-to-ground path
  loss for low altitude platforms in urban environments,'' in \emph{{Proc. IEEE
  Global Commun. Conf. (GLOBECOM)}}, 2014, pp. 2898--2904.

\bibitem{UAV_meas14}
J.~Holis and P.~Pechac, ``Elevation dependent shadowing model for mobile
  communications via high altitude platforms in built-up areas,'' \emph{{IEEE
  Trans. Ant. Propag.}}, vol.~56, no.~4, pp. 1078--1084, 2008.

\bibitem{UAV_sim4}
Q.~Feng, J.~McGeehan, E.~K. Tameh, and A.~R. Nix, ``Path loss models for
  air-to-ground radio channels in urban environments,'' in \emph{{Proc. IEEE
  Vehic Technol. Conf. (VTC)}}, vol.~6, 2006, pp. 2901--2905.

\bibitem{Schneckenburger_Eucap16_GroundMp}
N.~Schneckenburger, T.~Jost, D.~Shutin, and U.~C. Fiebig, ``Line of sight power
  variation in the air to ground channel,'' in \emph{Proc. EuCAP}, Davos,
  Switzerland, 2016.

\bibitem{UAV_sim8}
I.~J. Timmins and S.~O'Young, ``Marine communications channel modeling using
  the finite-difference time domain method,'' \emph{{IEEE Trans. Vehic.
  Technol.}}, vol.~58, no.~6, pp. 2626--2637, 2009.

\bibitem{UAV_sim12}
H.~Jamal, D.~W. Matolak, and R.~Sun, ``Comparison of {L-DACS} and {FBMC}
  performance in over-water air-ground channels,'' in \emph{{Proc. IEEE Digital
  Avionics Sys. Conf. (DASC)}}, 2015, pp. 2D6--1.

\bibitem{UAV_sim13}
H.~Jamal and D.~W. Matolak, ``{FBMC} and {LDACS} performance for future air to
  ground communication systems,'' \emph{{IEEE Trans. Vehic. Technol.}}, 2016.

\bibitem{UAV_geometric6}
S.~Blandino, F.~Kaltenberger, and M.~Feilen, ``Wireless channel simulator
  testbed for airborne receivers,'' in \emph{{Proc. IEEE Globecom Workshops (GC
  Wkshps)}}, 2015, pp. 1--6.

\bibitem{UAV_geometric5}
F.~Jiang and A.~L. Swindlehurst, ``Optimization of {UAV} heading for the
  ground-to-air uplink,'' \emph{{IEEE Journal on Selected Areas in Commun.}},
  vol.~30, no.~5, pp. 993--1005, 2012.

\bibitem{Optimum_height_UAV}
M.~M. Azari, F.~Rosas, K.~C. Chen, and S.~Pollin, ``Optimal {UAV} positioning
  for terrestrial-aerial communication in presence of fading,'' in \emph{Proc.
  IEEE Global Commun. Conf. (GLOBECOM)}, Dec. 2016, pp. 1--7.

\bibitem{raytracing}
Z.~Yun and M.~F. Iskander, ``Ray tracing for radio propagation modeling:
  principles and applications,'' \emph{IEEE Access}, vol.~3, pp. 1089--1100,
  2015.

\bibitem{UAV_sim16}
K.~Daniel, M.~Putzke, B.~Dusza, and C.~Wietfeld, ``Three dimensional channel
  characterization for low altitude aerial vehicles,'' in \emph{{Proc. Int.
  Symposium Wireless Commun. Sys. (ISWCS)}}, 2010, pp. 756--760.

\bibitem{UAV_sim18}
Y.~Wu, Z.~Gao, C.~Chen, L.~Huang, H.~P. Chiang, Y.~M. Huang, and H.~Sun, ``Ray
  tracing based wireless channel modeling over the sea surface near {D}iaoyu
  {I}slands,'' in \emph{{Proc. Int. Conf. Computational Intelligence Theory,
  Sys. and App. (CCITSA)}}, 2015, pp. 124--128.

\bibitem{Mr_Uwe1}
R.~Amorim, H.~Nguyen, P.~Mogensen, I.~Z. Kovács, J.~Wigard, and T.~B.
  Sørensen, ``Radio channel modeling for {UAV} communication over cellular
  networks,'' \emph{IEEE Wireless Commun. Letters}, vol.~6, no.~4, pp.
  514--517, Aug. 2017.

\bibitem{UAV_meas17}
M.~Simunek, P.~Pechac, and F.~P. Fontan, ``Excess loss model for low elevation
  links in urban areas for {UAV}s,'' \emph{Radioengineering}, 2011.

\bibitem{UAV_meas21}
T.~Tavares, P.~Sebastiao, N.~Souto, F.~J. Velez, F.~Cercas, M.~Ribeiro, and
  A.~Correia, ``Generalized {LUI} propagation model for {UAV}s communications
  using terrestrial cellular networks,'' in \emph{{Proc. IEEE Vehic. Technol.
  Conf. (VTC)}}, 2015, pp. 1--6.

\bibitem{Mr_Uwe2}
X.~Cai, A.~Gonzalez-Plaza, D.~Alonso, L.~Zhang, C.~B. Rodr{\'\i}guez, A.~P.
  Yuste, and X.~Yin, ``Low altitude {UAV} propagation channel modelling,'' in
  \emph{Proc. European Conf. on Ant. and Propag. (EUCAP)}.\hskip 1em plus 0.5em
  minus 0.4em\relax IEEE, 2017, pp. 1443--1447.

\bibitem{PL_FI}
\BIBentryALTinterwordspacing
{IST-4-027756 WINNER II }, ``{D1.1.2 V1.0 WINNER II} channel models,'' 2003,
  accessed: 2017-11-29. [Online]. Available:
  \url{http://www2.tu-ilmenau.de/nt/generic/paper_pdfs/Part%20II%20of%20D1.1.2.pdf}
\BIBentrySTDinterwordspacing

\bibitem{PL_DS}
T.~S. Rappaport \emph{et~al.}, \emph{Wireless communications: principles and
  practice}.\hskip 1em plus 0.5em minus 0.4em\relax Prentice hall PTR New
  Jersey, 1996, vol.~2.

\bibitem{AG_meas3}
J.~Child, ``Air-to-ground propagation at 900 {MH}z,'' in \emph{{Proc. IEEE
  Vehic. Technol. Conf. (VTC)}}, vol.~35, 1985, pp. 73--80.

\bibitem{UAV_sim15}
P.~J. Park, S.-M. Choi, D.~H. Lee, and B.-S. Lee, ``Performance of {UAV}
  (unmanned aerial vehicle) communication system adapting {WiBro} with array
  antenna,'' in \emph{{Proc. Int. Conf. Advanced Commun. Technol., (ICACT)}},
  vol.~2, 2009, pp. 1233--1237.

\bibitem{pathloss_new1}
M.~Mozaffari, W.~Saad, M.~Bennis, and M.~Debbah, ``Mobile unmanned aerial
  vehicles (uavs) for energy-efficient internet of things communications,''
  \emph{arXiv preprint arXiv:1703.05401}, 2017.

\bibitem{pathloss_new2}
D.~Athukoralage, I.~Guvenc, W.~Saad, and M.~Bennis, ``Regret based learning for
  uav assisted lte-u/wifi public safety networks,'' in \emph{Global
  Communications Conference (GLOBECOM), 2016 IEEE}.\hskip 1em plus 0.5em minus
  0.4em\relax IEEE, 2016, pp. 1--7.

\bibitem{pathloss_new3}
M.~Alzenad, A.~El-Keyi, and H.~Yanikomeroglu, ``3d placement of an unmanned
  aerial vehicle base station for maximum coverage of users with different qos
  requirements,'' \emph{IEEE Wireless Communications Letters}, 2017.

\bibitem{pathloss_new4}
E.~Kalantari, I.~Bor-Yaliniz, A.~Yongacoglu, and H.~Yanikomeroglu, ``User
  association and bandwidth allocation for terrestrial and aerial base stations
  with backhaul considerations,'' \emph{arXiv preprint arXiv:1709.07356}, 2017.

\bibitem{pathloss_new5}
R.~I. Bor-Yaliniz, A.~El-Keyi, and H.~Yanikomeroglu, ``Efficient 3-d placement
  of an aerial base station in next generation cellular networks,'' in
  \emph{Communications (ICC), 2016 IEEE International Conference on}.\hskip 1em
  plus 0.5em minus 0.4em\relax IEEE, 2016, pp. 1--5.

\bibitem{pathloss_new7}
M.~Alzenad, A.~El-Keyi, F.~Lagum, and H.~Yanikomeroglu, ``3d placement of an
  unmanned aerial vehicle base station (uav-bs) for energy-efficient maximal
  coverage,'' \emph{IEEE Wireless Communications Letters}, 2017.

\bibitem{PL_ITU}
\BIBentryALTinterwordspacing
{International Telecommunication Union}, ``Propagation data and prediction
  methods required for the design of terrestrial broadband millimetric radio
  access systems,'' 2003, accessed: 2017-11-27. [Online]. Available:
  \url{http://www.catr.cn/catr/catr/itu/itur/iturlist.jsp?docplace=P&vchar1=P.1410-2}
\BIBentrySTDinterwordspacing

\bibitem{AF_shadowing_Matolak}
R.~Sun, D.~W. Matolak, and W.~Rayess, ``Air-ground channel characterization for
  unmanned aircraft systems 8212;part {IV}: Airframe shadowing,'' \emph{IEEE
  Trans. Vehic. Technol.}, vol.~66, no.~9, pp. 7643--7652, Sept 2017.

\bibitem{BW_SSfading}
W.~Q. Malik, B.~Allen, and D.~J. Edwards, ``Impact of bandwidth on small-scale
  fade depth,'' in \emph{{Proc. IEEE Global Telecommuns. Conference
  (GLOBECOM)}}, Nov. 2007, pp. 3837--3841.

\bibitem{UAV_sim1}
Y.~Zheng, Y.~Wang, and F.~Meng, ``Modeling and simulation of pathloss and
  fading for air-ground link of {HAPs} within a network simulator,'' in
  \emph{{Proc. Int. Conf. Cyber-Enabled Distributed Computing and Knowledge
  Discovery }}, 2013, pp. 421--426.

\bibitem{MIMO_antenna}
F.~Bohagen, P.~Orten, and G.~E. Oien, ``Design of optimal high-rank
  line-of-sight {MIMO} channels,'' \emph{{IEEE Trans. Wireless Commun.}},
  vol.~6, no.~4, 2007.

\bibitem{UAV_sim_mimo1}
D.~Rieth, C.~Heller, D.~Blaschke, and G.~Ascheid, ``On the practicability of
  airborne {MIMO} communication,'' in \emph{{Proc. IEEE Digital Avionics Sys.
  Conf. (DASC)}}, 2015, pp. 2C1--1.

\bibitem{UAV_geometric3}
P.~Chandhar, D.~Danev, and E.~G. Larsson, ``Massive {MIMO} as enabler for
  communications with drone swarms,'' in \emph{Proc. Int. Conf. on Unmanned
  Aircraft Systems (ICUAS)}, Jun. 2016, pp. 347--354.

\bibitem{CNPC_standard}
\BIBentryALTinterwordspacing
{Radio Technical Commission for Aeronautics}, ``Command and control {(C2)} data
  link minimum operational performance standards,'' accessed: 2017-07-07.
  [Online]. Available:
  \url{https://global.ihs.com/doc_detail.cfm?document_name=RTCA%20DO-362&item_s_key=00694348}
\BIBentrySTDinterwordspacing

\end{thebibliography}


\end{document}